\documentclass[11pt]{article}
\usepackage[numbers,sort&compress]{natbib}
\usepackage{latexsym}
\usepackage{amssymb,amsmath}
\usepackage{amsthm}
\usepackage{graphicx}
\usepackage{sgame}
\usepackage{color}
\usepackage{xcolor}
\usepackage{float}
\usepackage{appendix}

\usepackage{pifont}% http://ctan.org/pkg/pifont

\usepackage[hidelinks]{hyperref}
\usepackage{empheq}
\usepackage{blkarray}
\usepackage{cancel}
\usepackage{breqn}
\usepackage{comment}
\usepackage{todonotes}

\usepackage{makecell}

\usepackage{setspace}

\usepackage{enumerate}

\numberwithin{equation}{section}

\newcommand{\ZZ}{\mathbb{Z}}

\newcommand{\ds}{\displaystyle}

\newcommand{\ol}{\overline}

\DeclareMathOperator{\argmin}{argmin}

\newcommand{\bbm}{\begin{bmatrix}}
\newcommand{\bpm}{\begin{pmatrix}}
\newcommand{\ebm}{\end{bmatrix}}
\newcommand{\epm}{\end{pmatrix}}

 \newcommand{\dsdel}[2]{\displaystyle\frac{\partial #1}{\partial #2}}

\newcommand{\dsddx}[2]{\displaystyle\frac{d #1}{d #2}}
\newcommand{\dsddt}[1]{\displaystyle\frac{d #1}{dt}}

\newtheorem*{assumption*}{\assumptionnumber}
\providecommand{\assumptionnumber}{}


\usepackage{authblk}

\title{Evolutionary Dynamics Within and Among Competing Groups}
\author[1,2]{Daniel B. Cooney}
\author[3]{Simon A. Levin}
\author[1,2,4]{Yoichiro Mori}
\author[1,2,4]{Joshua B. Plotkin}
\affil[1]{Department of Mathematics, University of Pennsylvania, Philadelphia, PA, USA}
\affil[2]{Center for Mathematical Biology, University of Pennsylvania, Philadelphia, PA, USA}
\affil[3]{Department of Ecology and Evolutionary Biology, Princeton University, Princeton, NJ, USA}
\affil[4]{Department of Biology, University of Pennsylvania, Philadelphia, PA, USA}

\begin{document}

\newtheorem{definition}{Definition}[section]
\newtheorem{theorem}{Theorem}[section]
\newtheorem{lemma}[theorem]{Lemma}
\newtheorem{corollary}[theorem]{Corollary}
\newtheorem{claim}[theorem]{Claim}
\newtheorem{fact}[theorem]{Fact}
\newtheorem{proposition}[theorem]{Proposition}
\newtheorem{remark}[theorem]{Remark}
\newtheorem{observation}[theorem]{Observation}
\newtheorem{example}[theorem]{Example}

\renewcommand{\baselinestretch}{1.1}

% to get nice proofs ...
\newcommand{\qedsymb}{\mbox{ }~\hfill~{\rule{2mm}{2mm}}}

\maketitle

%\section*{Abstract}
\begin{abstract}
\noindent
Biological and social systems are structured at multiple scales, and the incentives of individuals who interact in a group may diverge from the collective incentive of the group as a whole. Mechanisms to resolve this tension are responsible for profound transitions in evolutionary history, including the origin of cellular life, multi-cellular life, and even societies. Here we synthesize a growing literature that extends evolutionary game theory to describe multilevel evolutionary dynamics, using nested birth-death processes and partial differential equations to model natural selection acting on competition within and among groups of individuals.  We apply this theory to analyze how mechanisms known to promote cooperation within a single group -- including assortment, reciprocity, and population structure -- alter evolutionary outcomes in the presence of competition among groups. We find that population structures most conducive to cooperation in multi-scale systems may differ from those most conducive within a single group. Likewise, for competitive interactions with a continuous range of strategies we find that among-group selection may fail to produce socially optimal outcomes, but it can nonetheless produce second-best solutions that balance individual incentives to defect with the collective incentives for cooperation. We conclude by describing the broad applicability of multi-scale evolutionary models to problems ranging from the production of diffusible metabolites in microbes to the management of common-pool resources in human societies.
\end{abstract}

{\hypersetup{linkbordercolor=black, linkcolor = black}
\begin{spacing}{0.01}
\renewcommand{\baselinestretch}{0.1}\normalsize
\tableofcontents
\addtocontents{toc}{\protect\setcounter{tocdepth}{2}}
%\addtocontents{toc}{~\vspace{-3\baselineskip}}
\end{spacing}
%\clearpage
\singlespacing

\section*{Introduction}

Life is intrinsically social at all scales. Even at cellular and sub-cellular scales the incentives of a replicating individual may differ from those of their surrounding group, which presents a form of social dilemma. Evolutionary dynamics can operate at multiple scales of biological organization simultaneously, often creating tension between the fate of individuals and the groups in which they live. % 
Mechanisms to relieve this tension are widespread and varied, and they have been the source of much complexity in biological phenotypes and behavior, including even the origin of cellular life itself \cite{sagan1967origin,margulis2008acquiring, sagan1967origin, bell2019masterpiece}. Failure to relieve this tension often results in pathology, disease, or social dysfunction \cite{aktipis2015cancer,aktipis2020cheating,feeny1990tragedy,ostrom2008tragedy}.

Competition at multiple scales of organization is especially acute in biological systems on the cusp of major evolutionary transitions \cite{szathmary1995major,okasha2005multilevel}, when the interests of individuals are at odds with the group incentives to form cooperative collectives \cite{levin2010crossing,folse2010individual}. This tension arises, for example, in the trade-off between an individual gene's incentive for rapid replication versus the collective need to establish a balanced ensemble of genes for healthy cellular function -- a conflict that can be mediated by genetic linkage, that is, by the origin of chromosomes \cite{gabriel1960primitive,smith1993origin,szathmary1993evolution,szilagyi2020evolution,cooney2022pde}, hypercycles \cite{eigen1971selforganization}, or package replication \cite{bresch1980hypercycles,niesert1981origin} and stochastic correctors  \cite{szathmary1987group,grey1995re}. Multilevel competition continues to play a role in transitions to yet higher levels of biological organization, including the origins of multicellularity \cite{tarnita2013evolutionary,staps2019emergence,pichugin2018reproduction}, such as the formation of multicellular colonies in {\it Volvox} \cite{herron2008evolution} and slime molds \cite{bonner1944descriptive,bonner1970induction}. In complex differentiated organisms, the breakdown of multicellular cooperation to support cellular-level cheating behaviors can manifest as disease, such as cancer, that threatens an organ or organism \cite{aktipis2015cancer,pacheco2014ecology,zaidi2019bottleneck,basanta2008evolutionary,basanta2008game}. Multilevel competition is also found across vast scales of biological and social organization, in the evolution of so-called ``evolutionary individuals'' that constitute units of selection \cite{michod1997cooperation,michod1997transitions,michod2007evolution}, from cells to societies.

Biologists have developed a robust theoretical framework to study strategic interactions in populations. Evolutionary game theory \cite{smith1973logic,smith1982evolution,hofbauer1998evolutionary,sandholm2010population,weibull1997evolutionary} originated from economic game theory, but it allows the study of dynamics and equillibria achieved by competition and differential reproduction, rather than by cognition and rationality. Biologists have uncovered a variety of mechanisms that can help reconcile individual incentives with behaviors that are beneficial to an entire group -- such as assortative interactions, population structure, and reciprocity. These mechanisms can emerge by evolution among individuals possessing little or no cognitive capacity \cite{trivers1971evolution,queller1984kin,grafen1979hawk,smith1982evolution,boyd1989evolution,nowak2006five,ohtsuki2006leading,ohtsuki2006simple,taylor2007transforming,van2014pathways}. Even in a single group of reproducing individuals, these mechanisms can promote cooperation or coordination to achieve collective outcomes that are far better than the Nash equilibrium of the underlying game. Nonetheless, these mechanisms may fail to promote any cooperation unless their impacts on individual fitnesses are sufficiently strong \cite{taylor2007transforming}; and some mechanisms may only promote a cooperative equilibrium as a possible alternative to a stable, all-defector outcome \cite{taylor2007transforming,boyd1990group}. Furthermore, mathematical models of mechanisms promoting cooperation via individual-level selection do not account for the possibility and impact of competition among groups. %

Evolutionary game theory can be generalized to describe natural selection operating simultaneously at multiple levels of organization in a group-structured population, characterizing both competition among individuals within a group and competition among groups of individuals \cite{traulsen2005stochastic,traulsen2006evolution,traulsen2008analytical,bottcher2016promotion,luo2014unifying,luo2017scaling,simon2010dynamical,simon2013towards,simon2016group,cooney2019replicator,cooney2020analysis,cooney2019assortment,cooney2022pde,velleret2020individual}.% 
This generalization provides a framework for exploring the countervailing effects of the individual incentive to defect and the collective incentive to cooperate, highlighting the evolutionary tug-of-war between levels of selection. Multilevel evolutionary game theory is relevant to questions as microscopic as the origins of chromosomes and cellular life, and as macroscopic as the dynamics of cultural evolution, cooperation, and conflict in human societies \cite{luo2014unifying,luo2017scaling,simon2010dynamical,simon2012numerical,simon2013towards,simon2016group,mcloone2018stochasticity,cooney2019replicator,boyd1990group,henriques2019acculturation,cooney2022pde}. 

In this paper we synthesize recent developments in multiscale evolutionary game theory \cite{luo2014unifying,luo2017scaling,cooney2019replicator,cooney2020analysis,cooney2019assortment,cooney2022long}; and we systematically contrast the outcomes of multiscale competition to those that can be achieved through mechanisms operating within a single group. We find qualitatively different long-term outcomes under competition in a single group, even including mechanisms such as assortment or reciprocity, than we find for simultaneous competition within and among groups. In most cases, however, neither within-group mechanisms nor among-group competition are sufficient to achieve collective optimality. And so we also analyze the combined effects of within-group mechanisms such as assortment, reciprocity, and population structure with the effects of among-group competition. In most cases these two mechanisms operate synergistically, providing greater collective benefits than expected under independence. We find that some within-group mechanisms, such as networked population structure, have qualitatively different effects on evolutionary dynamics when combined with among-group competition. And, in a few cases, the combination of within-group mechanisms and among-group competition can achieve collectively optimal outcomes.

Our presentation is structured as follows. In Section \ref{sec:model} we introduce a general model of multilevel selection in evolutionary games, using a partial differential equation (PDE) to describe strategic compositions across many groups. We analyze the long-time behavior of this model, highlighting the strength of among-group competition required to support cooperation in steady-state, and characterizing the long-term average payoff achieved. Then, in Section \ref{sec:simplerules}, we incorporate assortment, reciprocity, and network structure within groups, and we determine the critical benefit-to-cost ratio required to sustain long-time cooperation when these within-group mechanisms operate in concert with multilevel selection. In Section \ref{sec:kregularmain} we focus on multilevel selection when within-group interactions occur along $k$-regular graphs, showing that cooperation is sometimes maximized when individuals have an intermediate number of network neighbors. Finally, we study a game in which payoffs depend on continuous levels of effort exerted by cooperators and defectors. We show how increasing cooperator effort increases both the collective incentive to cooperate and the individual incentive to defect, allowing us to find a ``second-best" effort level that provides the best possible long-time collective payoff for a fixed strength of among-group selection. We conclude by discussing how our analysis of the tug-of-war between competition among individuals and among groups may be applied to study social dilemmas across scales, from genes to societies.

\section{General Modeling Framework}
\label{sec:model}

\subsection{Game-Theoretic Interactions}

To study the dynamics of multilevel selection, we generalize models from evolutionary game theory. We consider groups of individuals who engage in interactions with members of their peer group. Individuals can play one of two possible strategies, which we generically denote cooperate ($C$) or defect ($D$). In general, within a group containing a fraction $x$ of cooperators, a cooperator has reproduction rate $\pi_C(x)$ and a defector has reproduction rate $\pi_D(x)$, for some arbitrary  %
functions $\pi_C$ and $\pi_D$ satisfying $\pi_D(x) > \pi_C(x)$. 

We describe the evolution of behavior via individual-level selection in a large population by using the replicator equation, which tracks the proportion $x$ of cooperators in the group over time, according to their relative rate of reproduction
\begin{equation} \label{eq:withinreplicatorgeneric}
  \begin{aligned}
  \dsddt{x} &= x \left( 1 - x\right) \left( \pi_C(x) - \pi_D(x) \right) 
  \end{aligned}  
\end{equation}
This replicator equation can be derived either from models of reproductive competition via natural selection or from models of social learning based on individual-based decision-making \cite{taylor1978evolutionary,hofbauer1998evolutionary,sandholm2010population,cressman2014replicator}.

Much of our analysis holds for arbitrary frequency-dependent birth rates for two types, $C$ and $D$. But to gain intuition about cooperation \textit{per se}, we will often focus on pairwise games in which a cooperator pays a cost $c > 0$ to confer a benefit $b > 0$ to their opponent, whereas a defector pays no cost and confers no benefit. % 
Assuming that two cooperators produce an additional payoff $d$ representing possible (positive or negative) synergy for mutual cooperation \cite{queller1984kin}, the outcomes of such a pairwise interaction are represented by the following payoff matrix

\begin{equation} \label{eq:bcdpayoffmatrix}
\begin{blockarray}{ccc}
& C & D \\
\begin{block}{c(cc)}
C & b - c + d & - c \\
D & b & 0 \\
\end{block}
\end{blockarray}.
\end{equation}
Payoffs to individuals are assumed to determine their rates of reproduction. 
When a group is composed of fractions $x$ cooperators and $1-x$ defectors, individual payoffs to cooperators and defectors, averaged over all pairwise interactions, are given by $\pi_C(x) = (b+d)x - c$ and $\pi_D(x) = bx$, while the average payoff of group members is given by $G(x) = \left(b - c \right)x + d x^2$.  %When $c - b < d < c$, t
The payoff matrix of Equation \eqref{eq:bcdpayoffmatrix} corresponds to a Prisoners' Dilemma (PD) if its entries have the following ranking
\[b > b - c + d > 0 > -c \]
\cite{nowak2006evolutionary}. This ranking occurs provided that the synergy parameter $d$ satisfy the condition $c - b < d < c$. From the condition $d < c$, we see that defection is a dominant strategy and that $\pi_D(x) > \pi_C(x)$ for any $x$ between $0$ and $1$. For PD games in which $d < 0$, the condition $c-b < d$ tells us that $b > c$, and we can use this to show that the composition of cooperators $x^*$ that maximizes average-payoff $G(x)$ is given by
\begin{equation} \label{eq:x*}
    x^* = \left\{
     \begin{array}{cr}
       1 & : d \geq -\ds\frac{b-c}{2}\\
       \ds\frac{b-c}{2d} & : d < -\ds\frac{b-c}{2}
     \end{array}
   \right. .
\end{equation}
Notably, intermediate levels of cooperation can maximize the collective outcome for the group when the synergy $d$ of mutual cooperation is sufficiently negative.

The corresponding replicator equation within a group is simply 
\begin{equation} \label{eq:withinreplicator}
 \begin{aligned}
  \dsddt{x} &= x \left( 1 - x\right) \left( \pi_C(x) - \pi_D(x) \right) \\ &= x(1-x)(dx-c) \end{aligned}  
\end{equation}
For PD games (in which $d < c$), this replicator equation will always result in decreasing levels of cooperation, with the group converging to the all-defector composition $x=0$ in the long-time limit. To study the dynamics of multilevel selection in evolutionary games, we will consider an analogue of the replicator equation that describes the effects of both within-group and among-group competition.

\subsection{A Nested Model of Multilevel Selection}

To study the evolution of strategic behavior at multiple levels of selection, we consider multiple groups, each internally engaged in the strategic evolution described above. In addition to within-group strategic change caused by the differential reproduction of individuals, we also imagine among-group dynamics that arise when one group decides to copy the strategic composition of a different group -- or, equivalently, one group ``dies" and another group ``reproduces".

Figure \ref{fig:multilevel_cartoon} illustrates our model of multilevel strategic evolution, which 
follows the approach introduced by Luo and coauthors \cite{luo2014unifying,van2014simple,luo2017scaling,luo2013probabilistic}. 
This figure provides illustration of example individual-level and group-level birth-death events. One panel depicts the birth of a defector and the death of a cooperator within a single group, and the other panel depicts the birth of a two-cooperator group and the corresponding death of a three-defector group. 
The population state changes over time according to a ball-and-urn process, where the balls represent a given group and the urns represent the possible number of cooperators within groups. Within-group and among-group competition events correspond to balls taking local and non-local jumps between urns (Figure \ref{fig:multilevel_cartoon}).

\begin{figure}[H]
    \centering
   \includegraphics[width = 0.48\textwidth, height = .45\textwidth]{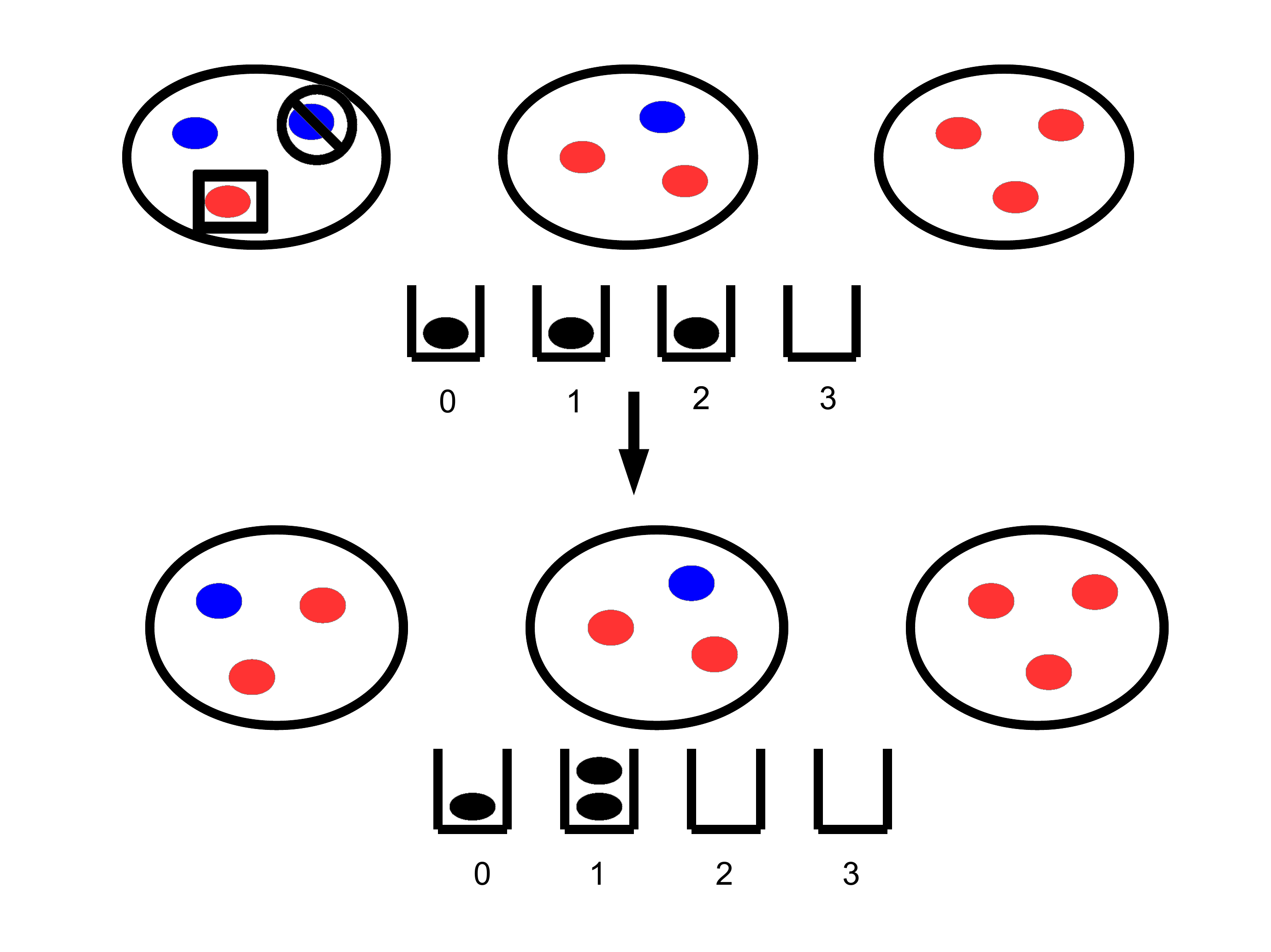}
   \includegraphics[width = 0.48\textwidth, height = .45\textwidth]{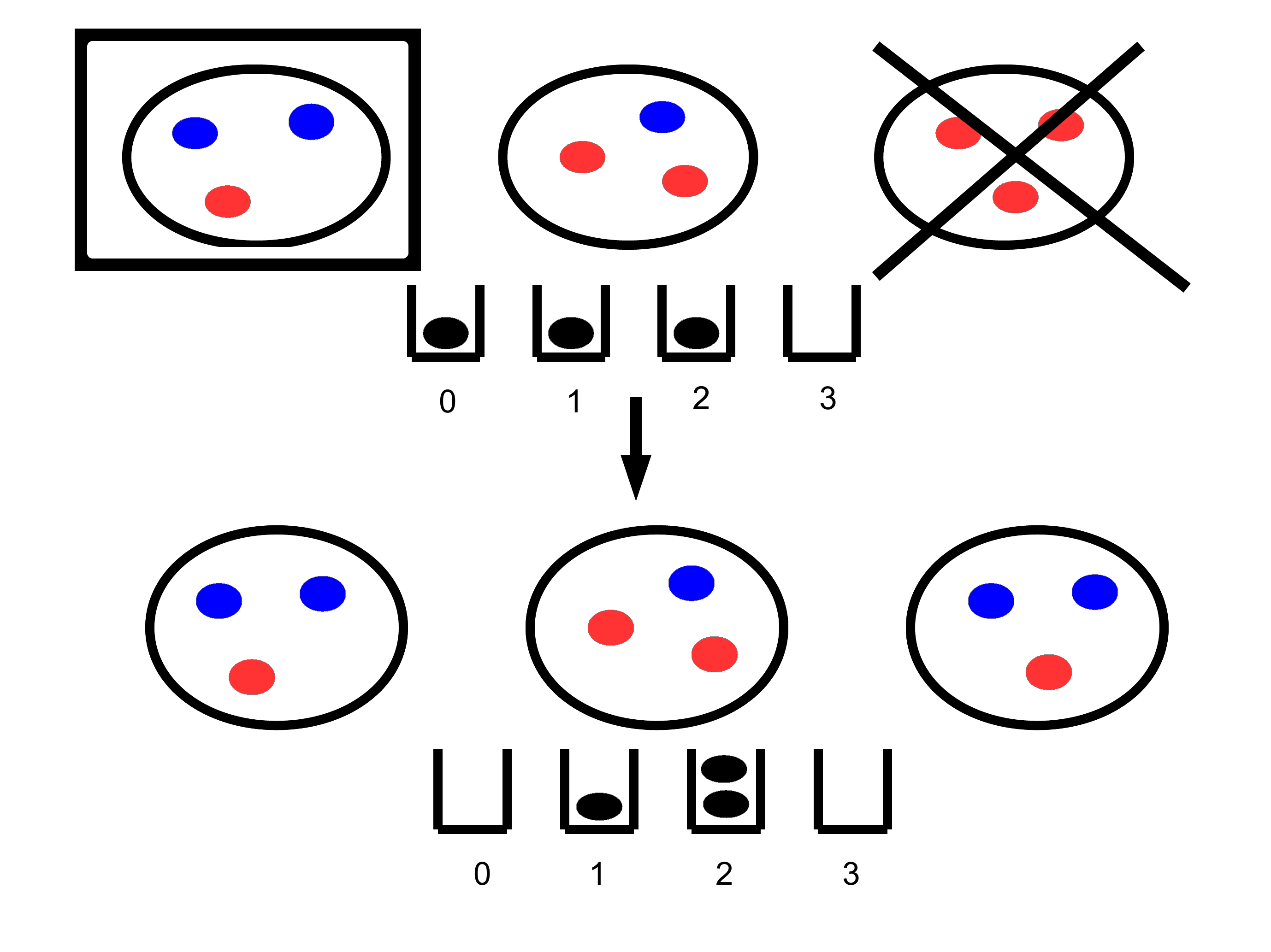}

    \caption[Schematic depiction of within- and among-group competition in population of three groups each comprised of three individuals.]{Schematic depiction of within- and among-group competition in population of three groups each comprised of three individuals. The left panel depicts an event in which a defector in the left-most group (square box) is chosen to reproduce, whose offspring replaces a cooperator in the same group (circled and crossed out). The urns describe the possible group compositions with labels between 0 and 3 representing the number of cooperators per group, and the balls in the urns correspond to one of the three groups. This individual-level birth and death event results in a local jump of the ball in the state space, as the middle group shifts from being a one-cooperator group to being a two-cooperator group. The right panel depicts an event in which a two-cooperator, one-defector group (shown in a square box) is chosen to reproduce, producing a copy of itself which replaces a three-defector group (which is crossed out), corresponding to a non-local jump in the state space. }
    \label{fig:multilevel_cartoon}
\end{figure}

\subsection{A PDE Limit for a Population of Many, Large Groups}

From this individual-based model of simultaneous competition within and among groups, we can derive a continuum description of the dynamics of multilevel selection in the limit of many of groups each of large size. The resulting partial differential equation (PDE) describes the evolution of the probability density $f(t,x)$ of groups containing a fraction $x$ cooperators at time $t$:

\begin{dmath} \label{eq:mainequation}
\dsdel{f(t,x)}{t} =  -  \overbrace{  \dsdel{}{x} \left( x(1-x) ( \pi_C(x) - \pi_D(x) )  f(t,x) \right)}^{\text{Within-Group Competition}}  + \lambda \underbrace{ f(t,x) \left[ G(x)  - \int_0^1 G(y) f(t,y) dy  \right]}_{\text{Among-group Competition}} 
\end{dmath}
where $\pi_C(x)$ and $\pi_D(x)$ are the individual payoffs of cooperators and defectors and $G(x)$ is the average payoff of group members in an $x$-cooperator group, and %
$\int_0^1 G(y) f(t,y) dy$ is the average payoff of the whole population. The first term in this equation corresponds to within-group competition based on individual payoffs, %
and it pushes each group towards the all-defector composition in the Prisoners' Dilemma (or towards a stable mix of cooperation and defection in other games, such as the Hawk-Dove game). The second term in Equation \ref{eq:mainequation} describes the impact of among-group competition, and it favors groups with high average payoff. The parameter $\lambda$ governs the relative importance of within-group and among-group competition, namely the ratio of selection strengths for within-group and among-group competition.
Equation \ref{eq:mainequation} is the multilevel analogue of the replicator equation presented in Equation \eqref{eq:withinreplicator}.   % 

Equation \eqref{eq:mainequation} is a first-order, non-local hyperbolic PDE, and so we can use the method of characteristics \cite{strauss2007partial,evans1998partial} to understand the multilevel evolutionary dynamics. %
The characteristic curves correspond to the traditional replicator equation for individual-level selection within each group, given by Equation \eqref{eq:withinreplicator}. And so the characteristic curves describe the impact of within-group selection, while the solutions along characteristics track the effect of among-group selection. 
While the derivation of Equation \eqref{eq:mainequation} is motivated by the payoffs generated by the game presented in Equation \eqref{eq:bcdpayoffmatrix}, we can also use Equation \eqref{eq:mainequation} to characterize the dynamics of multilevel selection for any continuously differentiable replication rates $\pi_C(x)$, $\pi_D(x)$, and $G(x)$ for which defectors have an individual-level advantage over cooperators ($\pi_D(x) > \pi_C(x)$) and all-cooperator groups have a collective advantage over all-defector groups ($G(1) > G(0)$).

The long-time behavior of solutions $f(t,x)$ to Equation \eqref{eq:mainequation} depends on the initial distribution of the strategic composition among groups. In particular, the dynamics depend on the density of the initial distribution near the full-cooperation equilibrium, $x=1$, as quantified by the H{\"o}lder exponent $\theta$ at $x=1$ \cite{luo2017scaling,cooney2022pde}.
Although the definition of the H{\"o}lder exponent $\theta$ is technical \cite{luo2017scaling,cooney2022pde}, we can think of $\theta$ as the inverse of the size of the initial cohort of high-cooperator groups in the population. (For example, the family of densities $f_{\theta}(x) = \theta \left( 1 - x\right)^{\theta - 1}$ have H{\"o}lder exponent $\theta$ near $x=1$.)

\subsection{The Long-Term Survival of Cooperation}
\label{sec:cooperationsurvival}
Selection within a group tends to promote individuals who defect in the prisoner's dilemma; where as selection among groups tend to promote cooperative groups. Provided the among-group selection is sufficiently strong compared to within-group selection, that is $\lambda$ exceeds some threshold $\lambda^*$, then cooperation will survive over the long-term. This result, and all the results summarized in this section, hold not only for pairwise games (Eq.~\ref{eq:bcdpayoffmatrix}), but also for arbitrary frequency-dependent reproduction rates satisfying $\pi_D(x) > \pi_C(x)$ and $G(1) > G(0)$.

In particular, when $\lambda > \lambda^*$, the composition $f(t,x)$ converges to a steady state density $f^{\lambda}_{\theta}(x)$ that supports positive levels of cooperation, and that depends upon the initial density according to its H{\"o}lder exponent $\theta$.
Conversely, when among-group selection is too weak, $\lambda \leq \lambda^*$, all groups become all-defectors in the long-time limit ($f(t,x)$ converges to the delta-function $\delta(x)$ concentrated at $x=0$).
The threshold selection strength that governs the longterm survival of cooperation is given by the following formula
\begin{equation} \label{eq:lambdastartintro} 
\lambda^* := \frac{( \overbrace{\pi_D(1) - \pi_C(1)}^{\substack{\text{Individual Incentive} \\ \text{to Defect}}}) \theta}{\underbrace{G(1) - G(0)}_{\substack{\text{Group Incentive} \\ \text{to Cooperate} }}}.
\end{equation}
We can interpret the numerator above as a defector's payoff advantage in a group otherwise composed of cooperators, while the denominator describes the collective advantage of being a full-cooperator group rather than a full-defector group. And so we see that the condition for cooperation to persist requires there be enough relative among-group selection to compensate for the temptation of an individual to defect in a a group of cooperators. In addition, note that the threshold selection strength is increasing in the H{\"o}lder exponent $\theta$, meaning that a population with a smaller initial cohort of groups near full-cooperation requires stronger among-group competition to sustain any long-term cooperation.

We can also explore how among-group competition improves collective outcomes by studying the average population payoff at steady state. When among-group selection is too weak, $\lambda \leq \lambda^*$, all groups converge to full defection, and the population average payoff is simply $G(0)$, the payoff of a full-defector group. But when group selection is sufficiently strong, $\lambda > \lambda^*$,  the mean population payoff converges to $G(x)$ averaged over the steady state density $f^{\lambda}_{\theta}(x)$ given by
\begin{equation}
    \langle G(\cdot) \rangle_{f^{\lambda}_{\theta}} = \int_0^1 G(x) f^{\lambda}_{\theta}(x) dx =  G(1) - \frac{\theta}{\lambda} \left(\pi_D(1) - \pi_C(1)\right). 
\end{equation}
And so in total, the long-term mean population payoff satisfies
\begin{equation} \label{eq:avgGfirstpiecewise}
     \lim_{t \to \infty}  \langle G(\cdot) \rangle_{f(t,x)} =
     \left\{
     \begin{array}{cr}
       G(0) & \text{when\ } \lambda \leq \lambda^* \\
        G(1)  - \frac{\theta}{\lambda} \left(\pi_D(1) - \pi_C(1)\right)& \text{when\ } \lambda > \lambda^*
     \end{array}
   \right. .
\end{equation}

Combining this formula with Equation \eqref{eq:lambdastartintro}, we can further understand the long-term collective outcome as 
\begin{equation} \label{eq:avgGlambstar}
     \lim_{t \to \infty}  \langle G(\cdot) \rangle_{f(t,x)} =  
     \left\{
     \begin{array}{cr}
       G(0) & \text{when\ } \lambda \leq \lambda^* \\
      \left[\frac{\lambda}{\lambda^*} \right] G(0) + \left[ 1 - \frac{\lambda^*}{\lambda}\right] G(1)& \text{when\ } \lambda > \lambda^*
     \end{array}
   \right. .
\end{equation}
This shows us that the collective outcome interpolates between the all-defector payoff $G(0)$, when $\lambda \leq \lambda^*$, to the all-cooperator payoff $G(1)$, as $\lambda \to \infty$. In particular, this means that the population can never achieve a mean payoff greater than in a group composed entirely of cooperators, which occurs in the limit of infinitely strong among-group competition. 
For games in which the full-cooperator group maximizes group payoff, strong competition among groups can bring the population towards optimality. However, if a group's payoff $G(x)$ is maximized by an intermediate proportion of cooperators, $x^*<1$, then the population can never achieve an optimal payoff, even for arbitrarily strong among-group competition.

We call this phenomenon the shadow of lower-level selection: for some games, no amount of among-group competition can erase the fact that a individual defector sees an advantage in a group with many cooperators. This phenomenon also manifests in the distribution of cooperators achieved at steady state. In the limit of infinitely strong among-group competition, the long-time population will concentrate at a group composition featuring a level of cooperation that is less than the social optimum and that achieves the same collective payoff as the all-cooperator group. In SI Section C, we provide a mathematical explanation for why the long-time collective payoff of the population is limited by that of the all-cooperator group.

Although this PDE model for multilevel selection considers infinitely many groups  of infinite size, the tradeoff between the collective incentive to cooperate and the individual incentive to defect also arise in finite populations. In SI Section \ref{sec:finite}, we consider a nested birth-death process introduced by Traulsen, Nowak, and coauthors \cite{traulsen2005stochastic,traulsen2006evolution,traulsen2008analytical} in a population with $m$ groups each composed of $n$ individuals, in which within-group competition acts on a faster timescale than among-group competition. In the case of games with the payoff matrix of Equation \eqref{eq:bcdpayoffmatrix}, we find that fixation of cooperation is favored over fixation of cooperation if the ratio $W$ of selection strengths for group-level and individual-level replication exceeds the threshold 
\begin{equation} \label{eq:Wstarmain}
W^* = \frac{n-1}{2 (m-2)} \left(\frac{\left( \pi_D(1) - \pi_C(1) \right) + \left(\pi_D(0) - \pi_C(0) \right) }{G(1) - G(0)} \right).
\end{equation}
Similar to our PDE model, this threshold selection strength depends on a ratio between the individual-level and group-level payoff comparisons. The relative difficulty of promoting cooperation also depends on the ratio $\frac{n}{m-2}$ between the size of groups and number of groups, %
which plays a role in moderating or amplifying the tug-of-war between individual and group-level incentives. 
%\textcolor{red}{Mention similarity between two models}

 \section{Simple Rules for the Evolution of Cooperation via Multilevel Selection.  } \label{sec:simplerules}

The results above show that the ability to support cooperation and achieve high collective payoffs via multilevel selection depends on the relative strength of within- versus among-group competition, $\lambda$, a defector's advantage in an otherwise cooperative group, $\pi_D(1) - \pi_C(1)$, the success of full-cooperator groups $G(1)$, and the composition of cooperators that maximizes a group's payoff, $x^*$. Because the ability to support cooperation via multilevel selection requires the collective incentive to cooperate to overcome the individual incentive to defect, we can explore biological mechanisms that promote cooperation either by improving the outcome of all-cooperator groups or by decreasing the individual-level temptation to defect among cooperators.  We summarize seven such mechanisms in Table \ref{tab:mechanismexplanation}, and we devote the remainder of this section to studying when and how these mechanisms work in concert with multilevel selection to promote the evolution of cooperation.

The seven mechanisms outlined in Table \ref{tab:mechanismexplanation} have already been studied in the context of a single level of selection (one group). Many of these mechanisms focus on ways to cluster cooperators with each other, or to incentivize cooperation in pairwise interactions. Two of these mechanisms model the effects of relatedness among individuals. The first is like-with-like assortment, in which individuals have a probability $r$ of interacting with individuals with the same strategy and probability $1-r$ of facing a randomly chosen individual, as relatedness \cite{grafen1979hawk} or assortative matching \cite{bergstrom2003algebra} can help pair cooperators with other cooperators. Another aspect of relatedness that can support cooperation is other-regarding preference, in which an individual's utility or reproductive potential depends not just on their own payoff (with weight 1), but also on the payoff of their interaction partners (with weight $F < 1$) \cite{taylor2007transforming,szabo2012selfishness}. We also consider mechanisms of reciprocity, in which cooperative individuals have a probability $q$ of detecting and punishing defectors based upon a reputation for defecting (indirect reciprocity), or individuals have repeated interactions with the same partner with discount rate $\delta$ and have the option of punishing defection in a given interaction by defecting in a future interaction (direct reciprocity). We also analyze the effect of network structure for strategic interactions in the case of $k$-regular graphs.
Finally, we consider a family of models in which payoff depends on the (continuous) level of effort expended by cooperators and defectors \cite{van2017hamilton}. In this model, cooperation featuring a higher level of effort increases the collective incentive to cooperate, but also introduces an increased incentive to defect against the more generous cooperators.  

Each of these mechanisms, which have been previously been studied in the context of a single level of selection, impacts the tug-of-war between individual-level selection for defectors and group-level selection for a high collective payoff. We summarize these effects of the mechanisms in Table \ref{tab:mechanisms}, highlighting the effect of each mechanism on (i) the individual-level incentive to defect in a group composed primarily of cooperators ($\pi_D(1) - \pi_C(1)$), (ii) the strategic composition that maximizes the average payoff of a group ($x^*$), and (iii) the collective incentive to have a full-cooperator group over a full-defector group ($G(1) - G(0)$). Some of these mechanisms, such as assortment or other-regarding preference, increase the individual-level incentive to defect in a cooperative group; whereas other mechanisms, such as birth-death dynamics on a network, decrease this incentive and other mechanisms, such as assortative matching have no effect on this incentive. Likewise, different mechanisms have different effects on the optimal level of cooperation within a group, as well as the collective benefit of pure cooperation.

\renewcommand{\arraystretch}{1.5}

\begin{table}[ht!]
\caption{Explanation of different mechanisms that may impact the tug-of-war between the collective incentive to cooperate and the individual incentive to cooperate.} %title of the table
 \label{tab:mechanismexplanation}
\centering % centering table
\begin{tabular}{|c|c|c|} % creating eight columns
\hline
Mechanisms  & %
\makecell{Explanation}%
& 
 %Increase $x^*$
 \makecell{References} 
\\

\hline

Assortment ($r$) & \makecell{Interact with same-strategy player with prob. $r$ \\ and random individual with prob. $1-r$}   & \cite{grafen1979hawk} \\
\hline
\makecell{Other-Regarding \\ Preference ($F$)} & \makecell{Place weight of $\frac{1}{1+F}$  own payoff \\ and weight $\frac{F}{1+F}$ on payoff of opponent} & \cite{taylor2007transforming}    \\

\hline
Indirect Reciprocity $(q)$ & \makecell{Cooperators identify defectors with probability $q$ \\ and punish by defecting}  &  \cite{nowak2005indirect}   \\
\hline
Direct Reciprocity & \makecell{Individuals defect against those who have \\ defected against them in prior interactions}  & \cite{axelrod1981evolution,trivers1971evolution}     \\
\hline
Network Reciprocity & \makecell{Individuals play game and update strategy\\  with local interactions on a $k$-regular graph}  &  \cite{ohtsuki2006simple,ohtsuki2006replicator}   \\
\hline

\hline
Continuous Levels of Effort & \makecell{Payoff from public good depends on the level \\ of effort characterizing cooperation and defection}   & \cite{van2017hamilton,iyer2020evolution}   \\
\hline
\end{tabular}
\end{table}

 \renewcommand{\arraystretch}{1.5}

\begin{table}[ht!] 
\caption{Impact of Mechanisms to Promote Cooperation via Multilevel Selection. Here we present the sign of the effect of each within-group mechanism on $\pi_D(1) - \pi_C(1)$, the individual-level incentive to defect, 
%in a group otherwise composed of cooperators, 
on $x^*$, the fraction of cooperators that maximizes group mean payoff, and on $G(1) - G(0)$, the collective incentive to achieve full-cooperation over full-defection. In the cases of $k$-regular graphs with death-birth or imitation updating, the sign of the effect on the individual incentive to defect, $\pi_D(1) - \pi_C(1)$, depends on the neighborhood size $k$ and the payoff matrix of the underlying game. Whereas all other mechanisms have a single, signed each of the key quantities that govern the outcome of multilevel selection.%
}
\label{tab:mechanisms}
\centering % centering table
\begin{tabular}{|c|c|c|c|} % creating eight columns
\hline
Mechanisms  & %$
\makecell{Effect on Individual \\ Incentive to Defect \\ $\pi_D(1) - \pi_C(1)$}%
& 
 %Increase $x^*$
 \makecell{Effect on Optimal \\Group  Composition \\ $x^*$} & 
\makecell{Effect on Collective \\ Incentive to Cooperate \\ $G(1) - G(0)$}\\
\hline
Assortment ($r$) & $\large{\boldsymbol{-}}$   & $\large{\boldsymbol{+}}$ & $\large{\boldsymbol{0}}$ \\
\hline
\makecell{Other-Regarding \\ Preference ($F$)} & $\large{\boldsymbol{-}}$  & $\large{\boldsymbol{0}}$  & $\large{\boldsymbol{0}}$   \\

\hline
Indirect Reciprocity $(q)$ & $\large{\boldsymbol{-}}$  &  $\large{\boldsymbol{+}}$   & $\large{\boldsymbol{0}}$  \\
\hline
Direct Reciprocity & $\large{\boldsymbol{+}}$  & $\large{\boldsymbol{+}}$   & $\large{\boldsymbol{+}}$   \\
\hline
\makecell{$k$-regular Graph \\
Death-Birth (DB) Updating} & $\large{\boldsymbol{+}}$ or  $\large{\boldsymbol{-}}$  &  $\large{\boldsymbol{+}}$   & $\large{\boldsymbol{0}}$  \\
\hline
\makecell{$k$-regular Graph \\ Birth-Death (BD) Updating} & $\large{\boldsymbol{+}}$  &  $\large{\boldsymbol{+}}$   & $\large{\boldsymbol{0}}$  \\
\hline
\makecell{$k$-regular Graph \\ Imitation (IM) Updating} & $\large{\boldsymbol{+}}$ or  $\large{\boldsymbol{-}}$  & $\large{\boldsymbol{+}}$   & $\large{\boldsymbol{0}}$  \\
\hline
\makecell{Level of  \\ Cooperator Effort ($e_C$)} & $\large{\boldsymbol{-}}$   & N/A    & $\large{\boldsymbol{+}}$   \\
\hline
\makecell{Level of  \\ Defector Effort ($e_D$)} & $\large{\boldsymbol{+}}$   & N/A    & $\large{\boldsymbol{-}}$   \\
\hline
\end{tabular}
\end{table}

Having summarized the qualitative effects of assortment and reciprocity in multilevel selection (Table \ref{tab:mechanismexplanation}), we now consider their quantitative effects within the broad class of three-parameter prisoner dilemma games (Equation \ref{eq:bcdpayoffmatrix}).
From Equation \ref{eq:lambdastartintro}, the threshold for achieving steady-state cooperation in such games is given by 
\begin{equation} \label{eq:pdthreshbcd}
\lambda^*_{PD} = \frac{\left(c-d\right) \theta}{b-c+d} = \left(\frac{b}{b-c+d} - 1\right) \theta,
\end{equation}
This tells us that the threshold selection level $\lambda^*_{PD}$ to support cooperation is a decreasing function of the synergy parameter $d$, so that increasing the synergy of mutual cooperation makes it easier to achieve cooperation via multilevel selection. 
 
  \renewcommand{\arraystretch}{1.8}

We can be more precise in our analysis of multilevel competition by focusing on a prisoner's dilemma with no synergy factor ($d=0$), which is called a simple donation game. We reformulate the threshold condition on $\lambda$ from Equation \ref{eq:lambdastartintro} by classifying games according to whether cooperation can or cannot survive under multilevel selection,  given $\lambda$ and $\theta$. To do this, we consider the donation game's payoff matrix
 \begin{equation} \label{eq:bcpayoffmatrix}
\begin{blockarray}{ccc}
& C & D \\
\begin{block}{c(cc)}
C & b - c & - c \\
D & b & 0  \\
\end{block}
\end{blockarray}
\end{equation}
where cooperators pay a cost $c$ to confer a benefit $b$ to their opponent, while defectors pay no cost and confer no benefit. We characterize such games by the minimum benefit-to-cost ratio $\frac{b}{c}$ for which cooperation is promoted by either individual or by multilevel selection. Table \ref{tab:simplerules} summarizes simple rules for the preservation of cooperation, showing how the critical benefit-to-cost ratio can differ for multilevel versus single-level competition, and how it depends on the parameter $\frac{\theta}{\lambda}$ of a multilevel process.
\renewcommand{\arraystretch}{2}
\begin{table}[H] 
\caption{Simple Rules for the Evolution of Cooperation via Multilevel Selection. Here, we present the threshold benefit-to-cost ratios $\frac{b}{c}$ required to produce cooperation for a single level of competition (single group), or under multilevel selection, for either well-mixed within-group interactions, as well as in-group dynamics that include a variety of different forms of assortment, reciprocity, or population structure. The last column indicates whether each mechanism of in-group structure facilitates cooperation across a broader class of games (lower $(b/c)^*)$ in the multilevel setting, compared to well-mixed interactions within each group -- that is, whether the in-group mechanism is synergistic with multilevel competition in promoting cooperation.} %title of the table
\label{tab:simplerules}
\centering % centering table
\begin{tabular}{|c|c|c|c|} % creating eight columns
\hline
Within-Group  & %
Threshold $\frac{b}{c}$  &  %
Threshold  $\frac{b}{c}$  & Mechanism Facilities \vspace{-2mm} \\

 Mechanism ($\varsigma$) &  (Individual) & (Multilevel) & Multilevel cooperation \\ 
\hline
Well-Mixed (WM) & $\infty$ & $1 + \ds\frac{\theta}{\lambda}$  & N/A \\
\hline
Assortment ($r$) & $\ds\frac{1}{r}$ & $1 + \ds\frac{ (1 - r) \theta }{\lambda + \theta r}$   & Always \\
\hline
Other-Regarding Preference ($F$) & $\ds\frac{1}{F}$  & $1 + \ds\frac{(1-F) \theta }{(1+F)\lambda  + F \theta}$  & Always \\
\hline
Indirect Reciprocity  $(q)$ & $ \ds\frac{1}{q} $ & $1 + \ds\frac{(1-q)\theta}{\lambda + \theta q} $  & Always \\
\hline
Direct Reciprocity $(\delta)$ & $ \ds\frac{1}{\delta} $ & $1 + \ds\frac{(1-\delta)\theta}{(1-\delta)\lambda +\delta \theta}$  & Always \\
\hline
$k$-regular Graph (DB) & $k$  & $1 + \ds\frac{k(k-1)\theta}{(k+1)(k-2) \lambda + k \theta}  $& $k > \ds\frac{2 \lambda}{\theta}$ \\
\hline
$k$-regular Graph (BD) & $\infty$ &$1 +  \ds\frac{k \theta}{(k-2) \lambda}$  & Never \\
\hline
$k$-regular Graph (IM) & $k+2$  & $1 + \ds\frac{(k+1) k \theta}{(k+3)(k-2) \lambda + k \theta}$ & $k > \ds\frac{6 \lambda}{\theta}$ \\
\hline
\end{tabular}
\end{table}
\renewcommand{\arraystretch}{1}

Table \ref{tab:simplerules} shows how large a range of donation games support cooperation via multilevel selection, under each of the seven assortment and reciprocity mechanisms we have studied.  For example, in the case of like-with-like assortment, the threshold benefit-to-cost ratio to support cooperation $\left(\frac{b}{c}\right)^*_{r,MS}$ is a decreasing function of the assortment probability $r$, because 
\begin{equation} \label{eq:bcstarrmultideriv}
    \dsdel{}{r} \left(\frac{b}{c}\right)^*_{r,MS} = - \frac{\theta \left( \lambda + \theta \right)}{\left( \lambda + \theta r \right)^2} < 0.
\end{equation}
This means that, as the assortment probability increases, the combination of multilevel selection and assortment helps to promote cooperation across a broader range of donation games. And so the presence of increasing assortment is always more conducive to cooperation under multilevel selection. By taking partial derivatives with respect to the other assortment and reciprocity parameters, we find that each such mechanism also has a monotonic effect -- always decreasing the critical benefit-to-cost ratio for cooperation.

\section{Multilevel Selection with Network Reciprocity: Non-monotonic Dependence on Neighborhood Size} \label{sec:kregularmain}

Compared to reciprocity and assortment, structured interactions along a network have a more nuanced impact on the evolution of cooperation in the multilevel setting. For a birth-death update rule in a structured population \cite{ohtsuki2006simple,ohtsuki2006replicator} cooperation is never favored in the single-level setting  \cite{ohtsuki2006simple,ohtsuki2006replicator}, regardless of the benefit-to-cost ratio, $b/c$. But cooperation can be supported in the multilevel setting for finite $b/c>1 + \frac{k \theta}{(k-2)\lambda}$. Nonetheless, this threshold is higher than the equivalent threshold for multilevel selection with well-mixed in-group interactions, $b/c>1 + \frac{\theta}{\lambda}$. And so graph structure with birth-death updating actually helps to inhibit the evolution of cooperation via multilevel selection.

Under either death-birth or imitation update rules \cite{ohtsuki2006simple,ohtsuki2006replicator}, the ability to support cooperation via multilevel selection depends on neighborhood size $k$. %
Figure \ref{fig:bcthresholdkregularDB} shows these effects in the case of death-birth updating (results are similar for imitation updating). For death-birth updating, the critical benefit-to-cost ratio required to achieve cooperation via multilevel selection exceeds the threshold when neighborhood size $k$ is sufficiently small ($k < \frac{2 \lambda}{\theta}$), while the threshold is less than the well-mixed case when individuals have sufficiently many neighbors ($k > \frac{2 \lambda}{\theta}$). By contrast, in a single group (single-level selection), cooperation is supported only if the benefit-to-cost ratio $b/c$ exceeds the network degree $k$ \cite{ohtsuki2006simple,ohtsuki2006replicator}, so the sparsest possible neighborhoods are most conducive to cooperation under individual-level selection. Consequently, placing interactions on a graph with suffciently small neighborhood size $k$ may help support support cooperation at the individual let, yet hurt support for cooperation via multilevel selection. Furthermore the critical benefit-to-cost ratio for multilevel selection for death-birth and imitation updating approach that of the baseline (well-mixed) multilevel selection model in the limit of infinite neighborhood size ($k \to \infty$). As a consequence, the dependence of multilevel selection on node degree $k$ is non-monotonic: the space of donation games that can support cooperation via multilevel selection may be maximized by an intermediate neighborhood size $k$ (see Figure 2, right).      %
As these results show, the within-group dynamics on $k$-regular graphs are not simply  endogenous versions of mechanisms of assortment and reciprocity, because they produce a variety of qualitatively different effects for cooperation via individual and multilevel selection.

\begin{figure}[ht!]
    \centering
    \includegraphics[width = 0.48\linewidth]{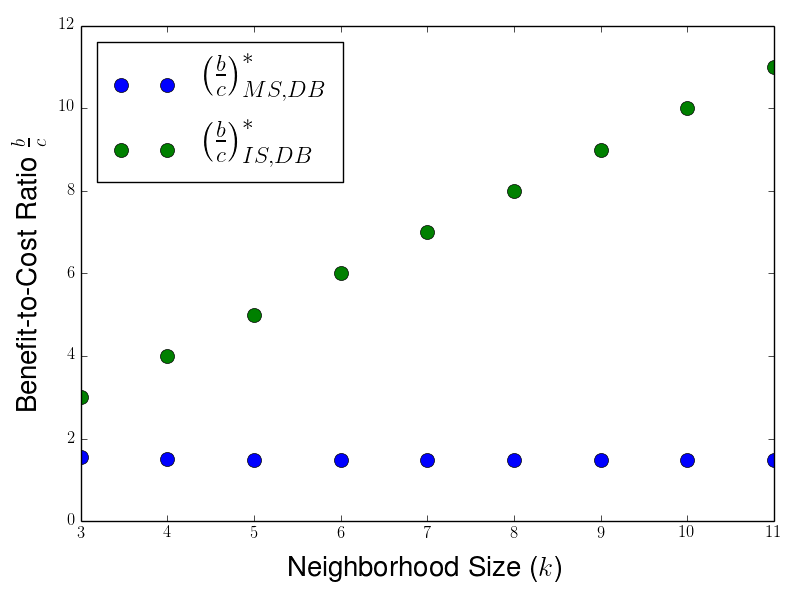}
    \includegraphics[width = 0.48\linewidth]{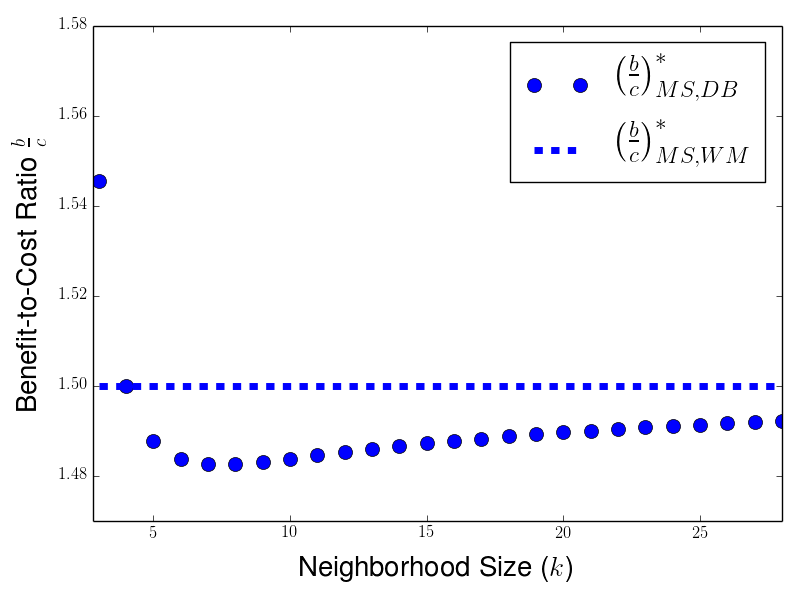}
    \caption{Plot of benefit-to-cost thresholds to support cooperation when within-group interactions take place on $k$-regular graphs with death-birth updating. Left: comparison between benefit-to-cost ratio required to support cooperation via multilevel selection $\left(\frac{b}{c}\right)^*_{IS,DB}$ (blue dots) and by individual-level selection alone $\left(\frac{b}{c}\right)^*_{IS,DB}$ (green dots). Right: comparison of threshold required to support cooperation by multilevel selection when within-group interactions take place on a $k$-regular graphs with death-birth updating $\left(\frac{b}{c}\right)^*_{MS,DB}$ (blue dots) and under well-mixed within-group interactions $\left(\frac{b}{c}\right)^*_{MS,WM}$ (blue dashed line).}
    \label{fig:bcthresholdkregularDB}
\end{figure}

In SI Section \ref{sec:graphs}, we extend our analysis of multilevel selection with graph-structured in-group interactions to consider a broader class of prisoner dilemma games, characterized by a payoff matrix (Equation \eqref{eq:bcdpayoffmatrix}) with negative synergy $d < 0$.  This class includes all PD games for which group average payoff is maximized by a mix of cooperators and defectors. 
We characterize the space of games for which graph-structured interactions on small $k$-regular graphs can either facilitate or impede cooperate via multilevel selection (SI Figures \ref{fig:DBparameter} and \ref{fig:IMparameter}). 
We also explore how the neighborhood size $k$ governs the threshold level of among-group competition $\lambda^*$ required for cooperation via multilevel selection, where again intermediate connectivity is optimal for cooperation (SI Proposition \ref{prop:klambdastarminDB}
and SI Figure \ref{fig:DBlambdathreshold}).

\section{A Model with Continuous Effort Levels}
\label{sec:continuouseffort}

To explore the tug-of-war between collective and individual incentives in a setting with richer detail, we study a family of social dilemmas in which the payoffs depend on a continuous levels of effort. The levels of effort characterizes the degree of cooperation or defection \cite{van2017hamilton} . In this model, an individual exerting effort $e$ who interacts with an opponent exerting effort $\tilde{e}$ receives payoff 
\begin{equation} \label{eq:uetildeemain}
    u(e,\tilde{e}) = a \tilde{e} - e \tilde{e}.
\end{equation}
The socially optimal level of effort producing greatest collective payoff is $e^*=\tfrac{a}{2}$, although evolution by individual-level selection will never achieve this optimal outcome. Nonetheless, as we will see, multilevel selection can approach this social optimum.

We consider a pair of effort levels $e_C>e_D$ expended by a cooperator and defector, so that the utility function of a two-player two-action game (Equation \eqref{eq:uetildeemain}) is described by the payoff matrix
\begin{equation} \label{eq:continuouspayoffparametrizedmain}
\begin{blockarray}{ccc}
& C & D \\
\begin{block}{c(cc)}
C & e_C (a - e_C) & e_D (a - e_C) \\
D & e_C (a - e_D) & e_D (a - e_D) , \\
\end{block}
\end{blockarray} \:.
\end{equation}
In a single population, evolution will always lead to full defection, which produces the lowest possible payoff. In fact, if effort is a mutable and heritable trait then evolution in a single population will always decrease effort towards zero \cite{geritz1998evolutionarily,diekmann2004beginner,brannstrom2013hitchhiker}. Here we study whether multilevel competition can support survival of the more cooperative effort level. A related question is whether the level of cooperation that multilevel selection supports can ever produce the maximal group payoff, which is achieved when $e=\tfrac{a}{2}$.

Given the pairwise payoffs above, the collective payoff in a full-cooperator and full-defector group are given by $G(1) = e_C (a - e_C)$ and $G(0) = e_D (a - e_D)$, while the individual payoffs near the all-cooperator composition are given by $\pi_C(1) = e_C (a - e_C)$ and $\pi_D(1) = e_C (a - e_D)$. In this continuous-effort setting, then, the threshold selection strength $\lambda^*(e_C,e_D)$ required to sustain some long-time cooperation (i.e.~some individuals contributing $e_C$) is given by 
\begin{equation} \label{eq:lambdastareCeD}
    \lambda^*(e_C,e_D) = \frac{\theta e_C}{a - e_C - e_D}.
\end{equation}
Note that $\lambda^*(e_C,e_D)$ is increasing in $e_C$ and $e_D$, so that increasing the effort level of cooperators or defectors requires stronger among-group selection to support cooperation in the long-run. 

%From Equations \eqref{eq:lambdastareCeD} and \eqref{eq:G1eClambinf}, 
The results above show that increasing the level of cooperator effort has two contrasting effects on cooperation.
Increasing the cooperator effort $e_C$ makes sustaining cooperator more difficult, especially when among-group competition is weak, because it increases the individual-level incentive to defect $\pi_D(1) - \pi_C(1)$ more than it affects the collective incentive to cooperate $G(1) - G(0)$. But when the cooperator effort is larger, this increases the collective outcome $G(1)$ for the all-cooperator group, and thereby increases the best possible outcome achievable via multilevel selection -- which is what matters when among-group competition is strong. We illustrate these two countervailing effects in Figure \ref{fig:effortlevelcollectiveoutcome}, where for low $\lambda$, populations with lower cooperator effort achieve greater long-time collective payoffs (Figure \ref{fig:effortlevelcollectiveoutcome}, left), while populations with higher levels of cooperator effort achieve better collective outcomes when among-group competition is strong.

The question remains what level of cooperator effort $e_C$ maximizes the average payoff at steady state, for a given strength of among-group competition. Provided that $\lambda>\lambda^*$, the average payoff is given by (see SI XX) 
\begin{equation}
    \langle G \rangle_{f^{\infty}} = e_C \left( a + \frac{\theta}{\lambda} e_D - \left( 1 + \frac{\theta}{\lambda} \right) e_C \right).
\end{equation}
In the limit as $\lambda \to \infty$ the collective payoff of the steady-state population is $e_C \left( a - e_C \right)$,
which is maximized for cooperator effort $e_C = \frac{a}{2}$. And so for sufficiently strong among-group competition, multilevel competition not only supports some cooperation, but it actually approaches the socially optimal level of cooperative effort.

However, when $\lambda$ is finite, the socially optimal effort $e_C = \tfrac{a}{2}$ does not, in fact, maximize average payoff under the dynamics of multilevel selection. Instead, we find that there is an effort level $\hat{e}_C<\tfrac{a}{2}$ that achieves the highest collective payoff under multilevel selection for given $\lambda$ (Figure \ref{fig:effortlevelcollectiveoutcome}, right). This level of cooperator effort can been seen as a second-best solution \cite{lipsey1956general,davis1965welfare,tilman2017maintaining} that arises from the dynamics of multilevel selection. The point is that increasing cooperative effort increases both the collective incentive to cooperate $G(1) - G(0)$ and the individual incentive to defect $\pi_D(1) - \pi_C(1)$; and the second-best effort level $\hat{e}_C$ provides the optimal balance of these opposing effects to promote collective payoff under multilevel competition.

In summary, our analysis shows that multilevel competition can support positive effort that would never evolve in a single group. However, even multilevel selection most favors an effort level that is second best to the socially optimal one, except in the limit of infinitely strong among-group competition.

\begin{figure}[H]
    \centering
    \includegraphics[width = 0.48\linewidth]{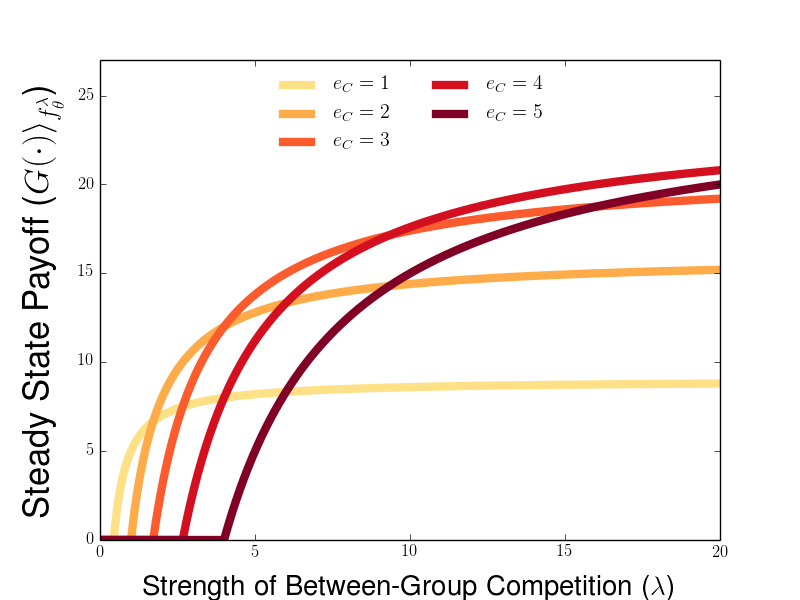}
     \includegraphics[width = 0.48\linewidth]{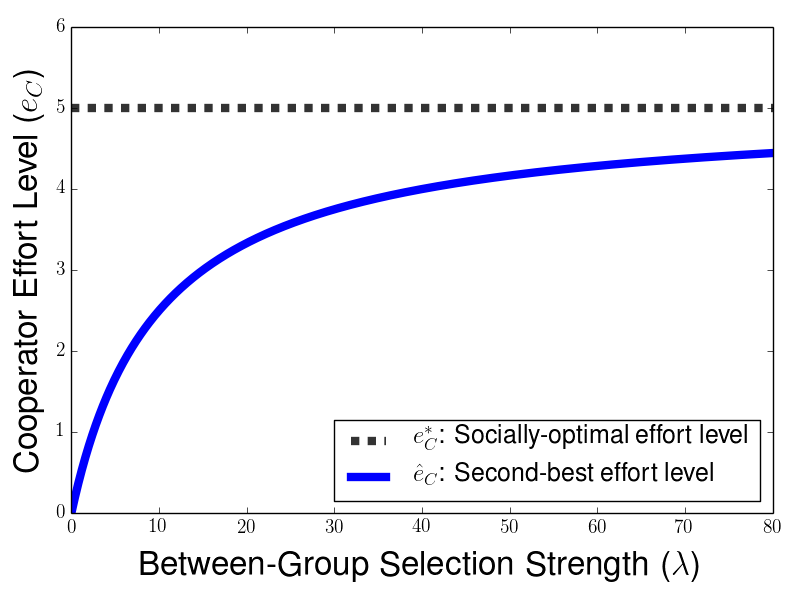}
    \caption{(left) Average payoff at steady state as a function of relative strength $\lambda$ of among-group competition, for different levels of cooperator effort $e_C$. 
    Defector effort is fixed at $e_D = 0$ and the socially optimal level of cooperator effort is fixed at $\tfrac{a}{2} = 5$. When the level of cooperator effect $e_C$ is larger, then stronger among-group competition $\lambda$ is required to support any cooperation in steady state. Nonetheless, for very strong among-group competition, large levels of cooperator effort increase the steady-state payoff, allowing the population to approach the socially optimal payoff, achieved by $e_C^*=\tfrac{a}{2}$, as $\lambda \to \infty$. (right) The "second-best" cooperation effort $e_C$ that maximizes long-term average fitness under multilevel selection, as a function of the strength of between group competition, $\lambda$. The second-best effort level approaches the socially optimal level $e_C^*=\tfrac{a}{2}$ as $\lambda \to \infty$.}
    \label{fig:effortlevelcollectiveoutcome}
\end{figure}

\section{Discussion}
\label{sec:discussion}

In this paper, we have synthesized  results from a general model for multilevel evolution, and systematically compared their outcomes to those that arise under single-level selection. In the context of a social dilemmas, we have derived ``simple rules" for how mechanisms of assortment and reciprocity, which are already known to facilitate cooperation within a single group, compare to and operative in combination with among-group competition. Our analysis highlights the key role of individual and average payoffs achieved at the all-cooperator and all-defector groups in determining the long-time support for cooperation.

The examples of individual-level and group-level replication we have studied highlight how tradeoffs between individual and collective incentives shape the long-time prospects for cooperation under multilevel selection.  In particular, we found that the mechanisms of assortment and reciprocity served only to decrease the individual-level incentive to defect, and had no impact on the average payoff of the all-cooperator group. This decreases the strength of among-group competition required to sustain long-time cooperation, but it does not allow for any improvement to collective payoff over the case of multilevel selection with well-mixed interactions in the limit of very strong between-group competition. By contrast, when within-group interactions take place on a $k$-regular graph, support for cooperation is maximized for graphs with an intermediate neighborhood size, in sharp contrast to the monotonic dependence on connectivity that is known to operate within a single group \cite{ohtsuki2006simple,ohtsuki2006replicator}.  As this result demonstrates, the PDE replicator equation provides additional insights into the interplay of social dilemmas and population structure, revealing more nuance on the collective benefits of cooperation than can been seen by studying individual-level selection on its own.

We have also analyzed a game in which payoffs to cooperators and defectors are parametrized by the (continuous) level of cooperator effort. In this case, unlike all others we have studied, there is a tradeoff between the collective incentive to cooperate and the individual incentive to defect, as we vary the effort expended by cooperators. For such games, we find that increased cooperator effort makes sustaining steady-state cooperation more difficult when among-group competition is weak, due to the increased individual-level advantage for defection; while at the same time the greater payoff to the all-cooperator group improve the collective outcome achieved when there is strong competition among groups. In this setting we can find second-best solutions for cooperator effort -- such that that an effort level that is socially suboptimal in a monomorphic population may still be the maximal collective payoff at steady state under multilevel selection. The key insight is that the ``second-best" solution balances in the individual incentive to defect versus the collective incentive to cooperate, when cooperation levels are continuous.

The flexibility of our PDE model for multilevel selection provides room for future applications of this framework. While we have focused primarily on contexts in which  payoffs arise from two-player, two-strategy symmetric matrix games, the method of analyzing the long-time behavior for our PDE carries through for any continuously-differentiable within-group and between-group replication rates that generalize the two-level evolutionary tension of the Prisoners' Dilemma. And so this framework can generate baseline predictions for the impact of competition within and among groups on the evolution on the evolution of cooperative traits or behaviors under a range of biological and social mechanisms, including, eg, public goods games.

Multilevel social dilemmas are widespread across biological and social systems, arising in settings such as the production of diffusible metabolic public goods \cite{de2013multilevel,nadell2010emergence,allen2013spatial}, the supply of energy to a cell by mitochondrial DNA \cite{gitschlag2020nutrient,rand2001units,haig2016intracellular}, replication control of plasmids \cite{paulsson2002multileveled}, and protocell evolution \cite{cooney2022pde}. Such conflicts also arise in infectious diseases, where they may manifest in the evolution of virulence when individual-level incentives for a pathogen to replicate are misaligned with collective incentives to spread between host organisms \cite{levin1981selection,dwyer1990simulation,coombs2007evaluating,gilchrist2004optimizing,gilchrist2006evolution,blackstone2020variation}. Explicit modeling of evolutionary dynamics operating at multiple levels of organization may be particularly relevant to empirical work on systems in which complementary genetic replicators compete at the individual level, such as in the case of heterotypic cooperation between viral strains \cite{xue2016cooperation} and genetic linkage of antibiotic resistance genes in bacteriophages \cite{sachs2005experimental},  %
or the tension between defective interfering particles and full viral genomes for within-host and between-host viral spread \cite{huang1970defective,manzoni2018defective,diaz2017sociovirology,rast2016conflicting}. The PDE models we have developed may also be relevant for studying cultural group selection, where simulation studies have been used to explore the evolution of altruistic punishment \cite{boyd2003evolution,janssen2014effect}, the evolution of social norms for indirect reciprocity \cite{santos2007multi}, and the role of institutions in promoting sustainable management of common-pool resources \cite{waring2017coevolution}.

Due to the wide range of biological and social scenarios in which selection operates at multiple scales, there are many directions for extending the approach we have developed in this study. Several variations on PDE models of multilevel selection have already been formulated, with examples including the multilevel dynamics with non-constant group size and group-level fission \cite{simon2016group,markvoort2014computer,traulsen2006evolution},  individual-level stochasticity even in the infinite population limit \cite{luo2017scaling,velleret2020individual}, multilevel competition with multiple types of groups that play different games \cite{cooney2022long}, and games with more than two strategies \cite{cooney2022pde}. Open areas for future research include developing PDE models to explore games with asymmetric roles for players, strategies for repeated games with large action spaces, pairwise competition between groups instead of global competition for collective replication, and  between-group population structure \cite{akdeniz2020cancellation}. Given this variety of future directions for mathematical development, there is still much to be learned from developing PDE models of evolutionary competition across scales.

\clearpage

\renewcommand{\abstractname}{Acknowledgments}
\begin{abstract} 
DBC received support from the National Science Foundation through grant DMS-1514606. DBC and YM received support from the Simons Foundation through the Math + X grant awarded to University of Pennsylvania. DBC and SAL received support from the Army Research Office through grant W911NF-18-1-0325. YM was supported by NSF DMS-2042144.

\end{abstract}

\bibliographystyle{unsrt}
\bibliography{references}

\appendix
\appendixpage
%\addappheadtotoc

In this appendix, we present complete derivations of formulas stated in the main text, and we provide additional analysis and discussion of our PDE model of multilevel selection. First, we provide a detailed presentation of our model of multilevel competition when payoffs are parametrized by continuous levels of effort of cooperators and defectors (Section \ref{sec:appendix-effort}). Next, we derive the simple rules for evolution of cooperation via multilevel selection, using the threshold group-level selection strengths for each within-group mechanism to obtain a critical benefit-to-cost ratio for multilevel selection in the simple donation game (Section \ref{sec:simplederivation}). Next we study the dynamics for our more general PDE model for multilevel selection, providing a mathematical illustration of the shadow of lower-level selection by showing that the long-time time-average of the population's collective payoff is limited by the payoff of all-cooperator groups (Section \ref{sec:longtimeillustration}). Next, we analyze the behavior of multilevel selection when within-group interactions take place on a $k$-regular graph, deriving threshold selection strengths as a function of the neighborhood size $k$ for three different update rules (Section \ref{sec:graphs}). Finally, we explore a model of multilevel selection in evolutionary games with finite population sizes, showing how the fixation probability of cooperation has a qualitative correspondence to our results for survival of cooperation in the infinite-population PDE limit (Section \ref{sec:finite}). 

\section{Two Strategy Games Parametrized by Levels of Effort}
\label{sec:appendix-effort}

In this section we consider two-strategy games where the payoffs generated by cooperators and defectors correspond to special cases of payoffs from social dilemmas in games with a continuous strategy space. We consider payoff functions $\pi(e,\tilde{e})$ received by a player using effort $e$ interacting with a peer with effort $\tilde{e}$. Then, if we define two distinguished levels of effort $e_C$ and $e_D < e_C$, then the interactions between players with these two possible strategies can be described by the following symmetric payoff matrix
\begin{equation} \label{eq:continuouspayoff}
\begin{blockarray}{ccc}
& C & D \\
\begin{block}{c(cc)}
C & u(e_C,e_C) & u(e_C,e_D) \\
D & u(e_D,e_C) & u(e_D,e_D) \\
\end{block}
\end{blockarray} \:.
\end{equation}

As an example, we can consider the following payoff function corresponding to a continuous-strategy Prisoners' Dilemma
\begin{equation} \label{eq:pieetilde}
    u(e,\tilde{e}) = a \tilde{e} - e \tilde{e},
\end{equation}
for $a > 0$. This game has been studied in the context of adaptive dynamics for the evolution of cooperation in the presence of assortative interactions \cite{van2017hamilton,iyer2020evolution}. For a monomorphic population with effort level $e$, each member of the population receives payoff $\pi(e,e) = e(a-e)$, and the socially-optimal level of effort is given by $e^* = \frac{a}{2}$. Considering two levels of effort corresponding to cooperation and defection, the payoff matrix from Equation \eqref{eq:continuouspayoff} now takes the following form
\begin{equation} \label{eq:continuouspayoffparametrized}
\begin{blockarray}{ccc}
& C & D \\
\begin{block}{c(cc)}
C & e_C (a - e_C) & e_D (a - e_C) \\
D & e_C (a - e_D) & e_D (a - e_D) , \\
\end{block}
\end{blockarray} \:.
\end{equation}
Now studying the multilevel dynamics of Equation \eqref{eq:mainequation} in terms of the payoff matrix of Equation \eqref{eq:continuouspayoffparametrized}, we see that the key quantities determining long-time behavior take the following form
\begin{subequations}
\begin{align}
G(1)  &= e_C (a - e_C) \\
G(0)  &= e_D (a - e_D) \\
\pi(1) &:= e_C (e_C - e_D).
\end{align}
\end{subequations}

Using these quantities and Equation \eqref{eq:lambdastartintro}, we can calculate the threshold level of among-group selection required to promote cooperation as a function of $e_C$ and $e_D$, which is given by
\begin{equation}
    \lambda^*(e_C,e_D) := \frac{\theta e_C}{a - e_C - e_D}.
\end{equation}
This tells us that the threshold needed to sustain cooperation is an increasing function of $e_C$, so more cooperative cooperators require a greater intensity of among-group competition in order to achieve cooperation via multilevel selection when faced with the same level of defector effort.

We can also use Equation \eqref{eq:avgGfirstpiecewise} to see that the long-time collective payoff achieved by the multilevel dynamics in a population playing the PD game from Equation \eqref{eq:continuouspayoffparametrized} is given by
\begin{equation}
\begin{aligned}
    \lim_{t \to \infty} \langle G(\cdot) \rangle_{f(t,x)}&= \max\left\{\lambda G(1) - \theta \pi(1), G(0) \right\} \\ &= \max\left\{\lambda \left[e_C (a - e_C) - x_D (a - x_D) \right] - \theta e_C (e_C - e_D), e_D (a - e_D) \right\}
    \end{aligned}.
\end{equation}
In the case of convergence to density steady states when $\lambda \left[G(1) - G(0) \right] > \theta \pi(1)$, we see that the long-time collective payoff has the following partial derivatives with respect to the effort levels of cooperators and defectors 
\begin{equation}
    \begin{aligned}
    \frac{\partial }{\partial e_C} \left(  \lim_{t \to \infty} \langle G(\cdot) \rangle_{f(t,x)} \right) &= \lambda (a - 2 e_C) - \theta (2 e_C - e_D) \\
     \frac{\partial }{\partial e_D} \left(  \lim_{t \to \infty} \langle G(\cdot) \rangle_{f(t,x)} \right) &=  \theta e_D.
    \end{aligned}
\end{equation}
Notably, this tells us that the long-time collective payoff is always increasing is the effort level of defectors, while, for a given defector effort level $e_D$ and given parameter values $\lambda$ and $\theta$, the long-time collective outcome is maximized by the following level of cooperator effort
\begin{equation}
    \hat{e}_C(e_D) := \frac{\lambda a + \theta e_D}{2 (\lambda + \theta)},
\end{equation}
which corresponds to a level of effort less than the socially-optimal level $e^* = \frac{a}{2}$ except in the limit of infinitely strong among-group competition as $\lambda \to \infty$. This level of cooperative effort can be thought of as a second-best solution that arises via multilevel selection under the circumstances of a fixed level of defector effort, which, for example, has been considered in continuous-strategy models for the efficient extraction of common-pool resources \cite{tilman2017maintaining}. Intuitively, we can understand the reason for this intermediate second-best option through the fact that increasing cooperator effort $e_C$ introduces a tradeoff between increasing the collective advantage of the full-cooperator group $G(1)$ and increasing the individual-level disadvantage $\pi(1)$ of cooperators relative to defectors in a group with many cooperators. The defector effort level $e_D$, however, only appears in the long-time average payoff through the individual level term $-\theta \pi(1)$, and therefore increasing $e_D$ simply decreases the individual incentive to cheat.

\section{Derivation of Simple Rules for Evolution of Cooperation via Multilevel Selection} \label{sec:simplederivation}

In this section, we will derive the critical benefit-to-cost ratios for the promotion of cooperation via multilevel selection both the baseline case of the donation game with well-mixed interactions and in concert with a variety of mechanisms of assortment and reciprocity. Here, we will present the known threshold selection strength $\lambda^*$ required to sustain cooperation for multilevel selection with a range of within-group mechanisms that have been derived in previous work, and show how these thresholds can be used to find the corresponding critical benefit-to-cost for multilevel selection in the donation game.

Previous work on Equation \eqref{eq:mainequation} in the context multilevel selection in evolutionary games has focused on the case of two-player, two-strategy social dilemmas \cite{cooney2019replicator,cooney2020analysis,cooney2019assortment}. The interactions in such games can be described by payoff matrices given by

\begin{equation} \label{eq:payoffmatrix}
\begin{blockarray}{ccc}
& C & D \\
\begin{block}{c(cc)}
C & R & S \\
D & T & P \\
\end{block}
\end{blockarray},
\end{equation}
where $R$ is the reward for mutual cooperation, $P$ is the punishment for mutual defection, $T$ is the temptation payoff received by defecting against a cooperator, and $S$ is the sucker payoff for cooperating with a defector. A game with the payoff matrix of Equation \eqref{eq:payoffmatrix} is called a Prisoners' Dilemma (PD) if the following ranking of payoffs is satisfied:
\begin{equation} \label{eq:PDranking}
    T > R > S > P.
\end{equation}

In a group featuring a fraction $x$ cooperators and a fraction $1-x$ defectors, the average payoff obtained by a cooperator and defector by playing each of its peer group members is given by
\begin{subequations} \label{eq:pifunctions}
\begin{align}
    \pi_C(x) &= Rx + (1-x) S \\
    \pi_D(x) &= Tx + (1-x) P.
\end{align}
\end{subequations}
The average payoff of group members in an $x$-cooperator is then given by
\begin{dmath} \label{eq:Gfunction}
G(x) = x \pi_C(x) + (1-x) \pi_D(x) = P + \left( S + T - 2P \right)x + \left( R - S - T + P \right) x^2. 
\end{dmath}
To simplify our description of how the within-group and among-group replication rates depend on the underlying payoff matrix of Equation \eqref{eq:payoffmatrix}, we introduce the following parameters:
\begin{subequations} \label{eq:payoffparameters}
\begin{align} 
    \alpha &= R - S - T + P \\
    \beta &= S - P \\
    \gamma &= S + T - 2P.
\end{align}
\end{subequations}
Using this paramterization, we can write the average payoff of the $x$-cooperator group as 
\begin{equation} \label{eq:Gabg}
    G(x) = P + \gamma x + \alpha x^2,
\end{equation}
and the individual-level advantage of defectors over cooperators in an $x$-cooperator group is given by
\begin{equation} \label{eq:pidiffabg}
    \pi_D(x) - \pi_C(x) = - \left(\beta + \alpha x\right).
\end{equation}
For these games with the payoff matrix of Equation \eqref{eq:payoffmatrix} and well-mixed interactions, the dynamics of multilevel selection follow a special case of Equation \eqref{eq:mainequation} given by
\begin{equation}
    \label{eq:PDEparam}
    \dsdel{f(t,x)}{t} = - \dsdel{}{x}\left[x(1-x)\left(\beta + \alpha x\right) \right] + \lambda f(t,x) \left[ \gamma x + \alpha x^2 - \int_0^1 \left( \gamma y + \alpha y^2\right) f(t,y) dy \right],
\end{equation}
and the dynamics of individual-level selection within groups is given by an ODE replicator equation of the form
\begin{equation} \label{eq:withinparam}
    \dsddt{x(t)} = x(1-x) \left( \beta + \alpha x \right).
\end{equation}
Using Equation \eqref{eq:lambdastartintro}, we can now see that, for such games, the relative strength of among-group selection required to sustain cooperation must satisfy
\begin{equation} \label{eq:wellmixedthreshold}
    \lambda > \lambda^* := \frac{\left(\pi_D(1) - \pi_C(1) \right) \theta}{G(1) - G(0)} = \frac{-\left( \beta + \alpha\right)
 \theta}{\gamma + \alpha}.
\end{equation}
For the case of the donation game with payoff matrix Equation \eqref{eq:bcpayoffmatrix}, we see that $\alpha = (b-c) - (-c) -(b) + 0 = 0$, $\beta = (-c) - 0 = -c$, and $\gamma = (-c) + (b) - 2(0) = b - c$. Plugging these expressions into Equation \eqref{eq:wellmixedthreshold}, we see that the relative selection strength to sustain cooperation for the multilevel dynamics in the donation game is
\begin{equation} \label{eq:lambdastarbc}
    \lambda > \lambda^* = \frac{c \theta}{b-c}.
\end{equation}
Furthermore, we can rearrange this inequality to derive the following benefit-to-cost threshold to sustain cooperation is
\begin{equation} \label{eq:bcthreshwellmixed}
    \frac{b}{c} > 1 + \frac{\theta}{\lambda}. 
\end{equation}
This threshold is increasing in the H{\"o}lder exponent $\theta$ near $x=1$, and decreasing in the relative selection strength $\lambda$ of among-group competition. 

In Table \ref{tab:lambdathresholdmechanisms}, we extend this analysis of threshold relative selection strengths and benefit-to-cost ratios models of multilevel selection in evolutionary games featuring interactions with a variety of within-group population structures. This table presents the threshold selection strengths for general PD games with well-mixed interactions \cite{cooney2020analysis}, in models with assortment, other-regarding preference and direct and indirect reciprocity derived in past work \cite{cooney2019assortment}, and for models with interactions on $k$-regular graphs derived in Section \ref{sec:graphlambdainfinity} of this paper. We then calculate the threshold relative selection strength for the special case of the payoff matrix for the donation game (as in Equation \eqref{eq:lambdastarbc}, which, in turn, allows us to derive the critical benefit-to-cost ratio to sustain long-time cooperation via multilevel selection (as in Equation \eqref{eq:bcthreshwellmixed}). The last column describing critical benefit-to-cost ratio is presented in Table \ref{tab:simplerules} to illustrate simple rules for the evolution of cooperation via multilevel selection in concert with a range of within-group population structures. 

\begin{table}[H] 
\caption{Threshold relative levels of among-group competition $\lambda$ and critical benefit-to-cost ratios required to achieve cooperation via multilevel selection across a range of within-group mechanisms for altering individual-level payoffs or interactions.} %title of the table
\label{tab:lambdathresholdmechanisms}
\centering % centering table
\begin{tabular}{|c|c|c|c|c|} % creating eight columns
\hline
\makecell{Within-Group \\ Interactions}  & %
\makecell{Threshold $\lambda$ : \\ General PD}%
& 
 \makecell{Threshold $\lambda$: \\ Donation Game} & %
\makecell{Threshold $\frac{b}{c}$ \\ Donation Game} %
\\

\hline
Well-Mixed \cite{cooney2020analysis} & $ \ds\frac{-(\beta + \alpha) \theta}{\gamma + \alpha}$ & $\ds\frac{c \theta}{b - c}$ & $1 + \ds\frac{\theta}{\lambda}$    \\
\hline
Assortment ($r$) \cite{cooney2019assortment} & $\ds\frac{\left[-(\beta+\alpha) - r\left( \gamma - \beta\right) \right]\theta}{\gamma + \alpha}$   & $\ds\frac{\left[c - rb\right] \theta}{b-c}$ & $1 + \ds\frac{(1-r)\theta}{\theta + \lambda}$  \\
\hline

\makecell{Other-Regarding \\ Preference \\ ($F$) \cite{cooney2019assortment}} & $\ds\frac{\left[ - \left(\beta + \alpha\right) - \left( \frac{F}{1+F} \right) \left( \gamma - 2 \beta \right) \right]\theta}{\gamma + \alpha}$  & $\ds\frac{\left[c - F b \right] \theta}{\left(1+F\right) \left( b-c \right)}$  & $1 + \ds\frac{(1-F) \theta}{(1+F)\lambda + F \theta}$   \\

\hline
\makecell{Indirect \\ Reciprocity \\ $(q)$ \cite{cooney2019assortment}} & $\ds\frac{\left[-\left( \beta + \alpha\right) - q \left(\gamma - \beta\right) \right] \theta}{\gamma + \alpha}$  &  $\ds\frac{\left[c- q b\right]\theta}{b-c}$   & $1 + \ds\frac{(1-q)\theta}{\lambda + q \theta}$   \\
\hline
\makecell{Direct \\ Reciprocity \cite{cooney2019assortment}} & $\ds\frac{\left[ - (\beta + \alpha) -  \delta (\gamma - \beta)  \right] \theta}{\gamma + \alpha}$  & $\ds\frac{\left[c - \delta b \right] \theta}{b-c}$   &  $1 + \ds\frac{(1-\delta)\theta}{\lambda + \delta \theta}$  \\
\hline
\makecell{$k$-regular \\ Graph (DB)} & $\frac{-\left[k \gamma + (k^2 - 1) \alpha + k(k-1) \beta \right]\theta}{(k+1)(k-2) \left(\gamma + \alpha\right)}$  &  $\frac{-\left[k(b-c) - k(k-1) c \right]\theta}{(k+1)(k-2) \left(b-c\right)}$  & $1 + \frac{k(k-1) \theta}{(k+1)(k-2) \lambda + k \theta}$   \\
\hline
\makecell{$k$-regular \\ Graph (BD)} & $\ds\frac{- \left[ (k-2)(\beta + \alpha )+ \alpha + 2 \beta) \right] \theta}{(k-2)(\gamma + \alpha)}$  &  $\left( \ds\frac{k}{k-2} \right) \ds\frac{c \theta}{b - c}$   & $1 + \left(\ds\frac{k}{k-2}\right) \ds\frac{\theta}{\lambda}$   \\
\hline
\makecell{$k$-regular \\ Graph (IM)} & $\frac{- \left[(k+3)(k-2)(\beta+ \alpha) + 3 (\alpha + 2 \beta) + k(\gamma + \alpha) \right] \theta}{(k+3)(k-2)(\gamma + \alpha)}$  & $\ds\frac{\left[ (k+2) c + b\right] k \theta}{(k+3)(k-2)(b-c)}$  & $1 + \frac{(k+1) k \theta}{(k+3)(k-2) \lambda + k \theta}$   \\
\hline
\end{tabular}
\end{table}

\section{Illustration of Multilevel Dynamics and Shadow of Lower-Level Selection}
\label{sec:longtimeillustration}

In this section, we analyze the dynamics and long-time behavior for our PDE model of multilevel selection. We show how to obtain solutions to our PDE by the method of characteristics, and we use the expressions for these solutions to provide justification and intuition for the shadow of lower-level selection. Namely, we will show how the long-time behavior of the time-averaged collective outcome of the population is limited by the collective outcome achieved by a single all-cooperator group. 

We consider a slight generation of our PDE model for the multilevel dynamics described by Equation \eqref{eq:mainequation}. Using the notation $\pi(x) = \pi_D(x) - \pi_C(x)$, we can rewrite Equation \eqref{eq:mainequation} in the form
\begin{dmath} \label{eq:PDEpiform}
 \dsdel{f(t,x)}{t} = \dsdel{}{x} \left[ x(1-x) \pi(x) f(t,x) \right] + \lambda f(t,x) \left[G(x) - \int_0^1 G(y) f(t,y) dy \right]. 
\end{dmath}
For this analysis, we will assume that $G(x)$ and $\pi(x)$ are continuously differentiable function on $[0,1]$, and consider multilevel competition characterized by $G(1) > G(0)$ and $\pi(x) > 0$ for $x \in [0,1]$. Biologically, these assumptions correspond to a generalization of the multilevel dynamics for Prisoners' Dilemmas games, encoding the properties that defectors have an individual-level advantage over cooperators and that an all-cooperator group has a collective advantage over an all-defector group. 

In Section \ref{sec:shadowilustration}, we use the solution for Equation \eqref{eq:PDEpiform} obtained through the method of characteristics to demonstrate long-time behavior consistent with the shadow of lower-level selection. For completeness, we present the derivation of the time-dependent solution to Equation \eqref{eq:PDEpiform} in Section \ref{sec:solutioncharacteristics}.

\subsection{Upper Bound on Mean Collective Fitness} \label{sec:shadowilustration}

Given an initial density of group compositions $f(0,x) = f_0(x)$, we can use the the method of characteristics to find the following solution to Equation \eqref{eq:PDEpiform}
\begin{equation} \label{eq:solutioncharillustrated}
f(t,x) = \underbrace{\exp\left( \lambda \int_0^t \left[G(\phi_s^{-1}(x)) - \langle G(\cdot) \rangle_{f(s,x)} \right] ds  \right)}_{\text{Between-Group Effects}} \underbrace{f_0(\phi_t^{-1}(x)) \dsdel{\phi_t^{-1}(x)}{x}}_{\text{Within-Group Effects}},
\end{equation}
where $\langle G(\cdot) \rangle_{f(t,x)} := \int_0^1 G(y) f(t,y) dy$ is the average group-level replication rate in the population and $\phi_t^{-1}(x)$ are the solutions to the backward characteristic curves that satisfy the ODE
\begin{equation} \label{eq:backwardchar}
    \dsddt{\phi_t^{-1}(x)} = \phi_t^{-1}(x) \left(1 - \phi_t^{-1}(x) \right) \pi\left( \phi_t^{-1}(x) \right) \: \: , \: \: \phi_0^{-1}(x) = x.
\end{equation}
We can obtain an upper bound on the collective outcome for the population by examining the effects of between-group competition. Using the change-of-variable $q = \phi_t^{-1}(x)$, we may compute that
\begin{dmath} \label{eq:integrandbounds}
    \exp\left( \lambda \int_0^t G(\phi_s^{-1}(x)) ds  \right) = \exp\left( \lambda \int_0^t \left[G(\phi_s^{-1}(x) - G(1) \right] ds \right) e^{\lambda G(1) t} = \exp\left( \lambda \int_x^{\phi_t^{-1}(x)} \frac{G(q) - G(1)}{q(1-q) \pi(q)} dq \right) e^{\lambda G(1) t} \leq
    \exp\left( \lambda \int_0^1 \frac{\left[G(q) - G(1) \right]_+}{q(1-q) \pi(q)} dq \right) e^{\lambda G(1)t},
\end{dmath}
where $[G(q)-G(1)]_+$ is the positive part of $G(q) - G(1)$, satisfying
\[ [G(q)-G(1)]_+ =  \left\{
     \begin{array}{cr}
       G(q) - G(1) & : G(q) > G(1)\\
       0 & : G(q) \leq G(1)
     \end{array}
   \right. .\]
For our general class of multilevel selection models, we have assumed $G(0) < G(1)$ and $G(x)$ is continuously differentiable on $[0,1]$. Under these assumptions the integrand on the right-hand side of Equation \eqref{eq:integrandbounds} is bounded as $q$ goes to $0$ and $1$. Therefore we see that there is a constant $C < \infty$ such that
 \begin{equation} \label{eq:G1bound}
     \exp\left( \lambda \int_0^t G(\phi_s^{-1}(x)) ds  \right) \leq C e^{\lambda G(1) t}.
 \end{equation}
We can view this bound as extracting $\lambda G(1)$ as a principal growth rate for solutions of the multilevel dynamics. Integrating both sides of Equation \eqref{eq:solutioncharillustrated} from $0$ to $1$, we can use the bound of Equation \eqref{eq:G1bound} to estimate that
 \begin{dmath}
     \int_0^1 f(t,x) dx = \int_0^1 \exp\left( \lambda \int_0^t \left[G(\phi_s^{-1}(x)) - \langle G(\cdot) \rangle_{f(s,x)} \right] ds  \right) f_0(\phi_t^{-1}(x)) \dsdel{\phi_t^{-1}(x)}{x} dx \leq
     C_1 e^{\lambda G(1) t} \exp\left(-\lambda \int_0^t \langle G(\cdot) \rangle_{f(s,x)}  ds  \right) \int_0^1 f_0(\phi_t^{-1}(x)) \dsdel{\phi_t^{-1}(x)}{x} dx  \leq C e^{\lambda G(1) t}  \exp\left(-\lambda \int_0^t \langle G(\cdot) \rangle_{f(s,x)}  ds  \right) \int_0^1 f_0(y) dy.
 \end{dmath}
Using the fact that the initial composition $f_0(x)$ and the solution $f(t,x)$ are probability densities, we have that $\int_0^1 f(t,x) dx = \int_0^1 f_0(y) dy = 1$ and can further deduce that
 \begin{equation} \label{eq:shadowmainbound}
    1 =  \int_0^1 f(t,x) dx \leq C_1 \exp\left( \lambda \left[G(1) t - \int_0^t \langle G(\cdot) \rangle_{f(s,x)}  ds  \right] \right).
 \end{equation}
If the time-average of the average group-level reproduction rate in the population satisfies
 \[ \frac{1}{t} \int_0^t \langle G(\cdot) \rangle_{f(s,x)}  ds > G(1) \]
for sufficiently large times $t$, then the right-hand side of Equation \eqref{eq:shadowmainbound} can be made arbitrarily small. This contradicts the fact that the solution $f(t,x)$ to the multilevel dynamics of Equation \eqref{eq:PDEpiform} is a probability density, and therefore we  conclude that
 \begin{equation}
     \label{eq:G1limsupbound}
     \limsup_{t \to \infty} \frac{1}{t} \int_0^t \langle G(\cdot) \rangle_{f(s,x)}  ds  \leq G(1).
 \end{equation}
In other words, the time-averaged group-level reproduction rate for the population cannot exceed the outcome of a single all-cooperator groups in the limit of large times $t$. 
 
The above calculations on the long-time time-averaged group-level reproduction rate provide some intuition for shadow of lower-level selection discussed in Section \ref{sec:cooperationsurvival}, showing the key role of  all-cooperator groups in maintaining cooperation via between-group cooperation. We can obtain more precise bounds on $\frac{1}{t} \int_t \langle G(\cdot) \rangle_{f(s,x)}  ds$ by exploring more refined estimates on the tail behavior of the initial density $f_0(\cdot)$ near the all-cooperator equilibrium. With the further assumption that the initial density $f_0(x)$ has well-defined H{\"o}lder exponent $\theta > 0$ near $x=1$, we can use our estimates of $f_0(\phi_t^{-1}(x))$ to further characterize the long-time value of $\langle G(\cdot) \rangle_{f(t,x)}  dt$ as given by Equation \eqref{eq:avgGfirstpiecewise} \cite{cooney2022long}. 

\subsection{Deriving Solutions via Method of Characteristics} \label{sec:solutioncharacteristics}

In this section, we show how to use the method of characteristics to derive the formula from Equation \eqref{eq:solutioncharillustrated}. To do this, we will generalize the approach the approach of Luo and Mattingly \cite{luo2017scaling} by first considering solutions to an associated linear PDE. Given a function $h(t)$ of time and using the shorthand $j(x) := -x (1-x) \pi(x)$, we first examine A PDE of the form
\begin{equation}
    \dsdel{f(t,x)}{t} = - \dsdel{}{x} \left[j(x) f(t,x) \right] + \lambda \left[G(x) - h(t) \right].
\end{equation}

To apply the method of characteristics, it helps for us to rewrite our equation in the following form:
\begin{dmath} \label{eq:PDEcharform}
 \dsdel{f(t,x)}{t} + j(x) \dsdel{f(t,x)}{x} = f(t,x) \left[\lambda  \left( G(x) - h(t) \right)  + j'(x) \right]. 
\end{dmath}
From this form of the equation, we see that the characteristic curves are given by solutions $\phi_t(x_0)$ to the ODE
\begin{equation} \label{eq:charODEgeneral}
    \dsddx{x(t)}{t} = j(x) = - x (1-x) \pi(x) \: \: , \: \: x(0) = x_0.
\end{equation}
We can then keep track of the solution of Equation \eqref{eq:PDEcharform} along a given characteristic curve $\phi_t(x_0)$ by studying $f(t,\phi_t(x))$. Differentiating this quantity with respect to time, we see that the density along characteristics satisfies
\begin{equation} \label{eq:solutionalongchar}
\begin{aligned}
\dsddx{}{t} f(t,\phi_t(x_0)) &= \dsdel{f(t,\phi_t(x_0))}{t}  + \dsdel{f(t,\phi_t(x_0))}{\phi_t(x_0)} \left[\dsddx{}{t} \phi_t(x_0) \right] \\ 
&= \dsdel{\rho(t,\phi_t(x_0))}{t}  
+ j(\phi_t(x_0)) \dsdel{f(t,\phi_t(x_0))}{\phi_t(x_0)} \\ &= f(t,\phi_t(x_0)) \left[ \lambda \left(G(\phi_t(x_0)) - h(t) \right) - j'(\phi_t(x_0)) \right]% 
\end{aligned}
\end{equation}
We can then integrate in time to see that solutions along characteristics satisfy
\begin{dmath}
    f(t,\phi_t(x_0)) = f(x_0) \exp\left( \int_0^t \left[ \lambda G(\phi_s(x_0)) - \lambda h(t) - j'(\phi_t(x_0)) \right] ds \right),
\end{dmath}
where $f_0(x) = f(0,x)$ is the initial density for our population. Using the substitution $u = \phi_s(x_0)$ and the fact that $\dsddx{u}{s} = \dsddx{}{s} \phi_s(x_0) = j(\phi_s(x_0))$, we can further see that
\begin{equation} \label{eq:solutionschangeofvariable}
    f(t,\phi_t(x_0)) = f_0(x_0) \exp\left( \int_{x_0}^{\phi_t(x_0)} \left[ \lambda \left(\frac{G(u) - h(t)}{j(u)} \right) - \frac{j'(u)}{j(u)} \right] du \right).
\end{equation}
For the second term in the integral, we see that
\[ - \int_{x_0}^{\phi_t(x_0)} \frac{j'(u)}{j(u)} du = - \log\left(j(u) \right) \bigg|_{x_0}^{\phi_t(x_0)} = \log\left( \frac{j(x_0)}{j(\phi_t(x_0))}\right)  \]
Applying this to Equation \eqref{eq:solutionschangeofvariable}, we can further see that
\begin{equation}
    f(t,\phi_t(x_0)) = f_0(x_0) \left( \frac{j(x_0)}{j(\phi_t(x_0))} \right) \exp\left( \int_{x_0}^{\phi_t(x_0)} \lambda \left[\frac{G(u) - h(t)}{j(u)} \right] du \right).
\end{equation}
This has given us an expression for $f(t,\phi_t(x_0))$, expressing our solution to the associated linear PDE along characteristic curves. We can make this an expression an explicit function of $x$ by using the backward characteristic curves given by $x_0 = \phi_t^{-1}(x)$. Substituting this into our expression above, we see that
\begin{equation} \label{eq:explicitlinear} f(t,x) = f_0(\phi_t^{-1}(x)) \left( \frac{j(\phi_t^{-1}(x))}{j(x)} \right) \exp\left( \int_{\phi_t^{-1}(x)}^{x} \lambda \left[\frac{G(u) - h(t)}{j(u)} \right] du \right). \end{equation}
From \cite[Lemma 3.3]{cooney2022long}, we further know that the backward characteristics $\phi_t^{-1}(x)$ satisfy the following equation
\begin{equation} \label{eq:backwardJacobian} \dsdel{\phi_t^{-1}(x)}{x} = \frac{\phi_t^{-1}(x) \left(1 - \phi_t^{-1}(x) \right) \pi\left( \phi_t^{-1}(x) \right)}{x (1-x) \pi(x)} = \frac{j(\phi_t^{-1}(x))}{j(x)}. \end{equation}
Next we plug this into Equation \eqref{eq:explicitlinear}. Combining this with the substitution $\phi_s^{-1}(x)$ for $u$ (and the corresponding fact that $\dsddx{}{s} \phi_s^{-1}(x) = \phi_s^{-1}(x) (1-\phi_s^{-1}(x)) \pi(\phi_s^{-1}(x)) = -j(\phi_s^{-1}(x))$), we can rewrite Equation \eqref{eq:explicitlinear} as
\begin{equation}
    f(t,x) = f_0(\phi_t^{-1}(x)) \dsdel{\phi_t^{-1}(x)}{x} \exp\left( \lambda \int_0^t \left[ G(\phi_s^{-1}(x)) - h(s) \right] ds  \right).
\end{equation}
Finally, we note that, by directly differentiating, we can check that this expression will be a solution to the nonlinear multilevel dynamics of Equation \eqref{eq:PDEpiform} if, and only if, $h(t) = \langle G(\cdot) \rangle_f(t,x) = \int_0^1 G(y) f(t,y) dy$. Therefore we see that the solution to Equation \eqref{eq:PDEpiform} with initial density $f_0(x)$ is given by the following expression:
\begin{equation} f(t,x) = f_0(\phi_t^{-1}(x)) \dsdel{\phi_t^{-1}(x)}{x} \exp\left( \lambda \int_0^t \left[ G(\phi_s^{-1}(x)) - \langle G(\cdot) \rangle_f(t,x)  \right] ds  \right), \end{equation}
as presented in Equation \eqref{eq:solutioncharillustrated}.

\section{Multilevel Selection with Interactions on $k$-regular Graphs}  \label{sec:graphs}

In this section, we explore the dynamics of multilevel selection when when the game-theoretic interactions and within-group selection events take place on a $k$-regular graph, a random graph where each node has exactly $k$ edges. For within-group dynamics, Ohtsuki and coauthors \cite{ohtsuki2006simple,ohtsuki2006replicator} considered the following three rules by which individuals.
\begin{itemize}
\item Death-Birth updating (DB): a random individual is chosen to die and its neighbors compete with probability proportional to payoff to reproduce and replace their neighbor 
\item Birth-Death updating (BD): an individual is chosen to reproduce with probability proportional to payoff, and their offspring replaces a randomly chosen neighbor.
\item Imitation updating (IM): an individual is chosen at random to possibly revise their strategy, and they pick either to imitate a neighbor's strategy or retain their own strategy with probability proportional to payoff
\end{itemize}

For each of these update rules, the authors derived effective replicator equations for the changing fraction of cooperators $x$ in a population living on $k$-regular graphs, playing a cooperative dilemma with the payoff matrix of Equation \ref{eq:payoffmatrix}. They derived the  so-called ``Ohtsuki-Nowak Transformation'', %
showing that the  change in strategic composition for individuals playing the game of Equation \eqref{eq:payoffmatrix} on the $k$-regular graph would follow a well-mixed replicator equation for the following transformed payoff matrix
\begin{equation} \label{eq:OhtsukiNowakpayoff}
\begin{blockarray}{ccc}
& C & D \\
\begin{block}{c(cc)}
C & R & S + b_k^{UR} \\
D & T - b_k^{UR} & P \\
\end{block}
\end{blockarray}, %
\end{equation}
where $b_k^{UR}$ is prescribed by the given update rule $UR \in \{DB, BD, IM \}$. These shifts in payoff were given in Equation 2 of Ohtsuki and Nowak \cite{ohtsuki2006replicator} by
\begin{subequations} \label{eq:bURformulas}
\begin{align} \label{eq:bDBformula}
    b_k^{DB} &= \frac{\left(k+1\right) R + S - T - \left(k+1\right) P}{\left(k+1 \right) \left(k - 2 \right)} = \frac{\alpha + 2 \beta + k \left(\gamma + \alpha \right)}{\left(k+1\right)\left(k-2\right)} \\
    b_k^{BD} &= \frac{R + S - T - P}{k-2} = \frac{\alpha + 2 \beta}{k-2} \label{eq:bBDformula} \\
    b_k^{IM} &= \frac{(k+3)R + 3S - 3T - (k+3) P}{(k+3)(k-2)} = \frac{3 \left( \alpha + 2 \beta\right) + k \left( \gamma + \alpha\right)  }{(k+3)(k-2)}.
\end{align}
\end{subequations}

Using the well-mixed individual-level replicator equation of Equation \eqref{eq:withinparam} for games with payoff matrices of the form of Equation \ref{eq:OhtsukiNowakpayoff}, the Ohtsuki-Nowak transformation tells us that the within-group replicator equations on $k$-regular graphs for update rule $UR$ are
\begin{equation}
    \begin{aligned}  \label{eq:graphwithingroupunnormalizedflux}
    \dsddt{x(t)} &= x \left( 1 - x \right)  j_k^{UR}(x)  \\
    j_k^{UR} &= \beta + b_k^{UR} + \alpha x.
    \end{aligned}
\end{equation}
The ODE has the usual endpoint equilibria of 0 and 1, as well as the potential interior equilibrium \begin{equation} \label{eq:xeqUR} x^{eq}_{UR} = - \left(\frac{\beta + b_k^{UR}}{\alpha}\right). \end{equation}

When interactions following a PD game take place on a $k$-regular graph, the all-defector equilibrium becomes unstable when $j_k^{UR}(0) > 0$, and cooperators outperform defectors for reproduction/imitation in groups with few cooperators. The full-cooperator equilibrium becomes stable under the within-group dynamics on graphs when $j_k^{UR}(1) > 0$, and cooperators outperform defectors for reproduction/imitation in groups with many cooperators. Furthermore, we can determine the possible bifurcation behaviors by examining the relative values of $j_k^{UR}(0)$ and $j_k^{UR}(1)$. 

Next, we consider the role that within-group graph structure plays in the competition among groups based on collective payoff. We first note that the average payoff achieved under well-mixed interactions for both the original payoff matrix of Equation \eqref{eq:payoffmatrix} and for the Ohtsuki-Nowak transformation of Equation \eqref{eq:OhtsukiNowakpayoff} are both given by $G(x) = P + \gamma x + \alpha x^2$. This is true because the main effect of transformed payoff matrix of Equation \eqref{eq:OhtsukiNowakpayoff} is to shift the payoffs of a cooperator-defector pair to $S + b_k^{UR}$ for the cooperator and $T - b_k^{UR}$, so such interactions still make contribution $\frac{T+S}{2}$ to the average payoff of group members.  %
As a consequence, we see that the payoff transformation does not actually reflect the changes in payoff to cooperators and defectors that occur when interactions are placed on a $k$-regular graph, but rather reflects the relative individual-level birth rates of cooperators and defectors without reference to payoffs. Although the transformed payoff matrix is generally useful for conceptually understanding and for empirical applications of the graph-structured dynamics in terms of ``effective game'' in the sense described by Kaznatcheev \cite{kaznatcheev2018effective,kaznatcheev2019fibroblasts,turner1999prisoner}, we will need to take an alternate approach to calculate the average payoff of group members to study the effects of among-group competition featuring within-group graph structure.

However, we can turn to the the pair approximations of Ohtsuki and coauthors \cite{ohtsuki2006simple,ohtsuki2006replicator} to determine the average payoff obtained in a population with fraction $x$ with interactions on a $k$-regular graph. In such a group, a cooperator and defector receive expected payoffs of
\begin{subequations} \label{eq:kpayoffneighbors} \begin{align}\pi_C^k(x) &= q_{C|C} R + q_{D|C} S = (1 - q_{D|C}) R + q_{D|C} S \\ \pi_D^k(x) &= q_{C|D} T + q_{D|D} P = q_{C|D} T + (1 - q_{C|D}) P, \end{align}
\end{subequations}
where $q_{A|B}$ is the conditional probability that the neighbor of an $B$ player is an $A$ player. Using the pair approximation of Ohtsuki and coauthors \cite{ohtsuki2006simple,ohtsuki2006replicator}, the conditional probability of connections reaches the equilibrium level
\begin{subequations}
\label{eq:pairprobs}
\begin{align} 
q_{C|D} &= \left( \frac{k-2}{k-1} \right) x \\
q_{D|C} &= \left(\frac{k-2}{k-1} \right) (1 - x)
\end{align}
\end{subequations}
for all three of our update rules. Plugging the interaction probabilities of Equation \ref{eq:pairprobs} into our expression for expected payoffs in Equation \ref{eq:kpayoffneighbors}, we further see that
\begin{subequations} \label{eq:pik}
\begin{align} \pi_C^k(x) &= R + \left( \frac{k-2}{k-1}\right) (S-R) (1-x)  \\  \pi_D^k(x) &= P + \left( \frac{k-2}{k-1}\right) (T-P) x \end{align}
\end{subequations}

%. 
Across all three update rules, we can therefore use Equation \eqref{eq:pik} to that the average payoff for the members of an $x$-cooperator group $G_k(x) = x \pi_C^k (x) + (1-x) \pi_D^k$ is given by \begin{equation} \label{eq:Gkxpayoff} G_k(x)   = P + \left( \frac{k-2}{k-1}\right) \left( (S+T-2P)x + (R-S-T+P) x^2 \right) + \left( \frac{1}{k-1} \right) \left( R - P \right) x  \end{equation} 
Using the parameters $\alpha$, $\beta$, and $\gamma$ from Equation \eqref{eq:payoffparameters}, we can rewrite $G_k(x)$ as 
\begin{dmath} \label{eq:Gkparam} G_k(x) = P +  \left( \gamma + \frac{\alpha}{k-1} \right) x + \left(\frac{k-2}{k-1}\right) \alpha x^2 \\ = \left(  \frac{k-2}{k-1}\right) \left(P + \gamma x + \alpha x^2\right) + \left( \frac{1}{k-1}\right) \left( P + \left(\gamma + \alpha\right)  x \right). \end{dmath}

From Equation \ref{eq:Gkparam}, we see that $G_k(x)$ is a convex combination of the well-mixed group payoff function $G(x) = P + \gamma x + \alpha x^2$ for our the PD game under well-mixed interactions and $P + (R-P)x$, which favors as much cooperation as possible. This group-level replication rate interpolates between $G(x)$ as $k \to \infty$ and a function placing equal weight on $G(x)$ and $P + (R-P)x$ or $k = 3$. As a result, we see that the collective reproduction rate never full forgets the group-level effects for the shape of $G(x)$ even for the most sparse within-group network connectivity. %

However, for many of our games of interest, placing interactions onto a sufficiently sparse $k$-regular graph does shift the average payoff function $G_k(x)$ to make the full-cooperator groups the most competitive in among-group competition. We illustrate this through the following example, consisting of a family of PD games with $\beta = \alpha = -1$, previously studied because the well-mixed within-group dynamics are exactly solvable \cite{cooney2019replicator}.

\begin{example} \label{ex:alphaminus1} For the case in which $\alpha = -1$, we see that $$G_k'(x) = \gamma - \frac{1}{k-1}  -2  \left(\frac{k-2}{k-1} \right) x = 0 \: \: \mathrm{when} \: \:   x = \frac{\gamma(k-1)}{2(k-2)} - \frac{1}{2 (k-2)}$$  So for $k = 3$ and $\gamma \in [\frac{3}{2},2)$, we see that $G_3(x)$ has its only critical point at $ \gamma - \frac{1}{2} \in [1,\frac{3}{2})$, so $G_3(x)$ is maximized at $x_3^* = 1$. For the same game parameters in the case of well-mixed interactions, we see that $G(x) = \lim_{k \to \infty} G_k(x)$ is maximized at $x^*_{\infty} = \frac{\gamma}{2} \in [\frac{3}{4},1)$. Therefore a game with such $\gamma$ and $\alpha$ has average group payoff maximized at an interior fraction of cooperators in $(0,1)$ with well-mixed interactions, but allowing interactions for the same game to take place on a $3$-regular random graph allows average group payoff to be maximized with full cooperation. If instead $\gamma \in (1,\frac{3}{2}]$, we have that $G_3(x)$ is maximized by $x^*_3 \in (\frac{1}{2},1)$, so it is also possible that group-level replication may still be maximized by intermediate levels of cooperation on a $3$-regular graph. %
\end{example} 

What this example illustrates is that, for a two-player game for which the group average payoff $G(x)$ is maximized at an interior strategy $x^* \in (0,1)$ for well-mixed groups, it is possible for average group payoff to be maximized at $x^*_{3} = 1$ when within-group interactions occur on regular $k$ graph for $k = 3$. %In other words, the presence of graph structure within groups can facilitate further promotion of cooperation in selection at the among-group level. 
In particular, by increasing the point at which average group payoff $G_k(x)$ is maximized, decreasing the degree of social interactions $k$ can facilitate greater abundance of groups with large fractions of cooperators as $\lambda \to \infty$, and, therefore choosing small enough $k$ can shift the multilevel selection dynamics %from failing to promote any cooperation as $\lambda \to \infty$ to supporting cooperation, as well as shifting
from a regime featuring a shadow of lower-level selection to one in which full-cooperation is achieved in the limit of infinite among-group selection strength.

Having characterized the within-group birth-death dynamics and the average payoff of group members for games played on $k$-regular graphs, we can now define our multilevel dynamics by coupling within-group and among-group competition.  
Plugging the within-group dynamics of Equation \eqref{eq:graphwithingroupunnormalizedflux} and the group-level replication rate from Equation \eqref{eq:Gkparam} into Equation \eqref{eq:mainequation}, we see that we can describe a multilevel selection in which within-group interaction on $k$-regular graphs with an update rule $UR \in \{DB,BD,IM\}$ using the following PDE
\begin{dmath} \label{eq:graphpdegeneral}
 \dsdel{f(t,x)}{t} = - \dsdel{}{x} \left[ x (1-x) j_k^{UR}(x) f(t,x) \right] + \lambda f(t,x)  \left[ G_k(x) - \int_0^1 G_k(y) f(t,y) dy \right] .
\end{dmath}
%
%\begin{remark} \label{rem:pairapproximationweak} We 
We note that the pair approximations for $q_{C|D}$ and $q_{D|C}$ and within-group dynamics of Equation \ref{eq:graphwithingroupunnormalizedflux} derived by Ohtsuki and coauthors hold under certain weak-selection assumptions for individual-level competition \cite{ohtsuki2006simple,ohtsuki2006replicator}, so a rigorous derivation of Equation \eqref{eq:graphpdegeneral} from a two-level birth-death process with graph structure within-groups may require a careful analysis of the relative selection rates at the two levels.%

When full-defection dominates for a given update rule and neighborhood size $k$, we see from Equation \ref{eq:lambdastartintro} that the relative strength of among-group competition need  to achieve cooperation via multilevel selection must satisfy
\begin{equation} \label{eq:lambdastark}
\lambda > \lambda^*_{k,UR} =  \frac{-j_k^{UR}(1) \theta}{G_k(1) - G_k(0)}
\end{equation}
and we can see from Equation \ref{eq:avgGlambstar} that the average payoff of the population at steady state is given by
\begin{equation} \label{eq:Gklambdastark}
\langle G(\cdot) \rangle_{f^{\infty}(x)} = G_k(1) \left[1 - \frac{\lambda^*_{k,UR}}{\lambda} \right].
\end{equation}

When a population plays PD games with $\alpha < 0$ with a neighborhood size $k$ such that the within-group dynamics support a globally-stable equilibrium $x_{k,UR}^{eq}$, the multilevel dynamics of Equation \eqref{eq:graphpdegeneral} will resemble the generalized Hawk-Dove dynamics studied by Cooney and Mori \cite{cooney2022long}. % 
We can apply \cite[Equation C.12]{cooney2022long} to see that relative strength of among-group competition $\lambda$ needed to sustain steady-state cooperation in excess of $x^{eq}_{k,UR}$ must satisfy \begin{equation} \label{eq:lambdathresholdinteriork}
     \lambda > \lambda^{**}_{k,UR} := \frac{- j_k^{UR}(1) \theta}{G_k(1) - G_k\left(x^{eq}_{k,UR}\right)}.
 \end{equation}
 Furthermore, we can use \cite[Equation C.12]{cooney2022long} to see that the average payoff at density-steady states for $\lambda \geq \lambda^{**}_{k,UR}$ is given by 
 \begin{equation} \label{eq:steadystatepayofflambdastarstar}
 \langle G(\cdot) \rangle_{f(\cdot)} = \left(\frac{\lambda^{**}}{\lambda} \right)G_k(x^{eq}_{k,UR}) + \left(1 - \frac{\lambda^{**}}{\lambda} \right)  G_k(1).  
 \end{equation}

In the subsequent sub-sections, we will proceed to study the effects of $k$-regular graph structure and the different update rules on the ability to support cooperation through within-group selection and through multilevel selection. We will quantify the ability for graph dynamics to support cooperation through inequalities for the payoff parameters $x^* = \frac{\gamma}{-2\alpha}$ and $\frac{\beta}{\alpha}$, with an emphasis on understanding for which games can a given neighborhood size $k$ and update rule help a population achieve greater cooperation than is possible for well-mixed through individual and multilevel selection. We will discuss death-birth updating in Section \ref{sec:DB}, birth-death updating in Section \ref{sec:BD}, and imitation updating in Section \ref{sec:IM}.

\subsection{Death-Birth (DB) Updating} \label{sec:DB}

In this section, we will characterize the behavior of the multilevel dynamics for the DB update rule. We will focus on PD games in which $\alpha < 0$, including the subclass of PDs with intermediate payoff optima (which satisfy $\gamma + 2\alpha < 0$). In Proposition \ref{prop:dbthresholds}, we characterize conditions on the payoff parameters  $x^* := \frac{\gamma}{-2 \alpha}$ and $\frac{\beta}{\alpha}$ for which placing interactions on a $k$-regular graph with DB updating can help to faciliate cooperation via individual-level or multilevel selection. %
In particular, we find threshold levels the maximizer of the group payoff function $x^*$ to achieve these results beneficial to cooperation, which allows us to see whether graph reciprocity can help to produce greater cooperation when the underlying game-theoretic optimum favors groups with an intermediate level of cooperation.

\begin{proposition} \label{prop:dbthresholds}
Consider PD games with $\alpha < 0$ and within-group dynamics on a $k$-regular graph with DB updating. For such games, there exist the threshold quantities
\begin{subequations} \label{eq:dbthresholds}
\begin{align}
    T_0^{DB}(k) & :=  \frac{k+1}{2k} + \left( \frac{k-1}{2} \right) \frac{\beta}{\alpha} \\
    T_1^{DB}(k) & := \frac{(k-1)(k+1)}{2k} + \left(\frac{k-1}{2} \right) \frac{\beta}{\alpha} \\
     T_{\lambda^*}^{DB}(k) & :=  \frac{k+1}{2k} + \left( \frac{1}{k} \right) \frac{\beta}{\alpha}
\end{align}
such that the full-defector equilibrium is unstable when $x^* > T_0^{DB}(k)$, the full-cooperator equilibrium is stable when $x^* > T_1^{DB}(k)$, and multilevel selection is favored for $k$-regular graph interactions relative to well-mixed interactions (i.e. $\lambda^*_{k,DB} < \lambda^*_{k \to \infty} := \frac{-(\beta + \alpha) \theta}{\gamma + \alpha}$) when $x^* > T_{\lambda^*}^{DB}(k)$. Furthermore, these threshold quantities satisfy the following ranking:
\begin{displaymath}
 T_{\lambda^*}^{DB}(k) \leq T_0^{DB}(k) \leq T_1^{DB}(k) 
\end{displaymath}
\end{subequations}
\end{proposition}

The fact that there are games for which $x^* < T_{\lambda^*}^{DB}(k)$ means that introducing graph-structured interactions can hinder the achievement of cooperation of multilevel selection relative to the case of well-mixed game-theoretic interactions. This stands in contrast with the behavior of models of multilevel selection featuring assortment reciprocity, and other-regarding preference, in which adding greater chance of interacting with a same-strategy individual, greater weight on caring about one's opponents payoff, or the probability of a cooperator punishing a defector never harmed the steady state collective payoff or level of cooperation \cite{cooney2019assortment}.

\begin{proof}
We know that the full-defector equilibrium become unstable when
\begin{subequations} \label{eq:dbderivationequations}
\[ j_k^{DB}(0) = \beta + \frac{\alpha + 2 \beta + k (\gamma + \alpha)}{(k+1)(k-2)} > 0 \] Using the fact that $\alpha < 0$, we can rearrange this inequality to find that \begin{equation} \label{eq:xstardbT0} x^* := \frac{\gamma}{-2 \alpha} > \frac{k+1}{2} + \left(\frac{k-1}{2}\right) \frac{\beta}{\alpha}. \end{equation}
We can similarly derive the condition for stability of the full-cooperator equilibrium by knowing that it is stable when
\[ j_k^{DB}(1) = \beta + \alpha + \frac{\alpha + 2 \beta + k (\gamma + \alpha)}{(k+1)(k-2)} > 0 \]
and rearranging to find that 
\begin{equation}
 x^* := \frac{\gamma}{-2 \alpha} > \frac{(k-1)(k+1)}{2k} + \left(\frac{k-1}{2} \right) \frac{\beta}{\alpha}
\end{equation}
To determine whether the threshold $\lambda^*_{k,DB}$ to achieve steady state cooperation under multilevel selection is less than the threshold for multilevel selection in populations with well-mixed interactions, we can use Equation \ref{eq:lambdastark} and our expression for $\lambda^*$ to see that this requires
\begin{equation}  \frac{-j_k^{DB}(1) \theta}{G_k(1) - G_k(0)} <  \frac{- \left(\beta + \alpha\right) \theta}{G(1) - G(0)} \end{equation}
Noting that $G_k(1) = \gamma + \alpha = G(1)$ and $G_k(0) = 0 = G(0)$, we see that it now suffices to show that 
\[ j_k^{DB}(1) := \beta + \alpha + \frac{\alpha + 2 \beta + k (\gamma + \alpha)}{(k+1)(k-2)} > \beta + \alpha, \] 
and we can rearrange this inequality to show that $\lambda^*_{k,DB} < \lambda^*$ when
\begin{equation}
    x^* = \frac{\gamma}{-2 \alpha} > T_{\lambda^*}^{DB}(k) := \frac{k+1}{2} + \left(\frac{1}{k} \right) \frac{\beta}{\alpha}.
\end{equation}
\end{subequations}
Now we move on the the ranking of the thresholds. To show that $T_{\lambda^*}^{DB}(k) < T_0^{DB}(k)$, we use the fact that $\alpha,\beta < 0$ and $k \geq 3$ to see that 
\[ T_0^{DB}(k) - T_{\lambda^*}^{DB}(k) = \left(\frac{k-1}{2} - \frac{1}{k} \right) \frac{\beta}{\alpha} = \left( \frac{\left(k-2\right) \left(k+1\right)}{2k} \right) \frac{\beta}{\alpha} > 0\]
Similarly, to show that $T_1^{DB}(k) > T_{0}^{DB}(k)$, we use that $\alpha,\beta < 0$ and that $k \geq 3$ to see that  
\[T_1^{DB}(k) - T_{0}^{DB}(k) > \frac{(k-1)(k+1)}{2k} - \frac{k+1}{2k} = \frac{(k-2)(k+1)}{2k} > 0  \qedhere \]
\end{proof}

In Figure \ref{fig:DBparameter}, we illustrate the regions of the payoff parameter space for which within-group and among-group selection are hindered or favored by placing interactions on a $3$-regular graph relative to well-mixed interactions. From the thresholds in Equation \ref{eq:dbthresholds}, we are able to characterize these different qualitative regimes through the parameters $x^* = \frac{\gamma}{-2\alpha}$ and $\frac{\beta}{\alpha}$. Going from left to right, the four regions depicted are
\begin{itemize}
    \item games that do not satisfy the requirements to be a PD game (as $x^* < \frac{1}{2}$ so $\gamma + \alpha < 0$ and $R < P$)
    
    \item games for which $x^* < T_{\lambda^*}^{DB}(3)$, so within-group selection cannot produce cooperation and the threshold to produce cooperation via multilevel selection is higher than the case of well-mixed interactions 
    
    \item games for which $ T_{\lambda^*}^{DB}(3) < x^* <  T_{0}^{DB}(3)$, so within-group selection can't produce cooperation on its own but the threshold is decreased for promotion cooperation via multilevel selection
    
    \item games for which $x^* > T_{0}^{DB}(3)$, so cooperation is facilitated both for within-group selection and for multilevel selection
\end{itemize} 
Additionally, we notice that all three qualitative regimes for the PD game occupy some portion of the parameter space in which $x^* < 1$. This means that, for games with in which intermediate levels of cooperation maximize collective payoff, situating interactions on a 3-regular graph can either help or hurt the achievement of cooperation via individual and multilevel selection relative to the scenario of well-mixed game-theoretic interactions.

 \begin{figure}[ht!]
     \centering
     \includegraphics[width = 0.75\textwidth]{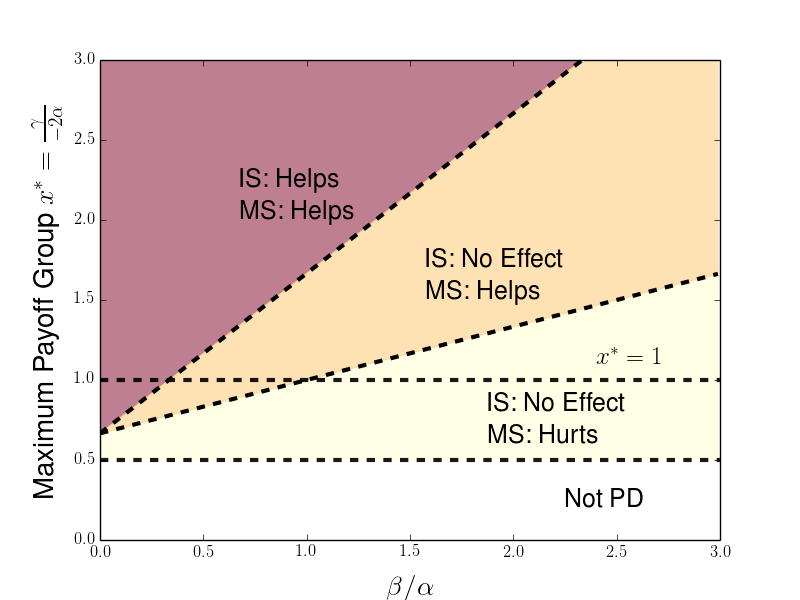}
     \caption[Regions of payoff parameter space in which multilevel selection and individual-level selection is helped or harmed by placing interactions on a $k$-regular graph with DB updating, relative to well-mixed social interactions.]{Regions of payoff parameter space in which multilevel selection and individual-level selection is helped or harmed by placing interactions on a $3$-regular graph, relative to well-mixed social interactions. Payoff matrices parametrized by values of group optimum $x^*$ and the ratio $\frac{\beta}{\alpha}$. Top panel provides a parameter regime including both edge and intermediate group optima, while right panel focuses on the case of intermediate group optima. IS and MS stand for individual-level selection and multilevel selection, respectively.}
     \label{fig:DBparameter}
 \end{figure}

 We further address the issue of threshold relative selection strength $\lambda^*_{k,DB}$ in Figure \ref{fig:DBlambdathreshold}. Here we plot this threshold quantity as a function of $k$ for three different values of $\beta$ (dots), and compare these to the threshold selection levels for well-mixed interactions (dashed lines). For $\beta = -1$ (black dots) and $\beta = -\frac{1}{2}$, we see that there are sufficient small neighborhood sizes $k$ for which the threshold is increased by placing interactions on the $k$-regular graph, while for $\beta = -\frac{1}{5}$ we see that the threshold quantities $\lambda^*_{k,DB}$ are less the threshold $\lambda^*_{PD}$ in the well-mixed case for all neighborhood sizes $k$. For all three cases, we see that there are intermediate values of $k$ that minimize the threshold quantity $\lambda^*_{k,DB}$, telling us that intermediate sparsity of interactions can be most favorable for trying to establish cooperation via multilevel selection for games with intermediate payoff optima $x^* < 1$. %

  \begin{figure}[ht!]
     \centering
     \includegraphics[width = 0.75\textwidth]{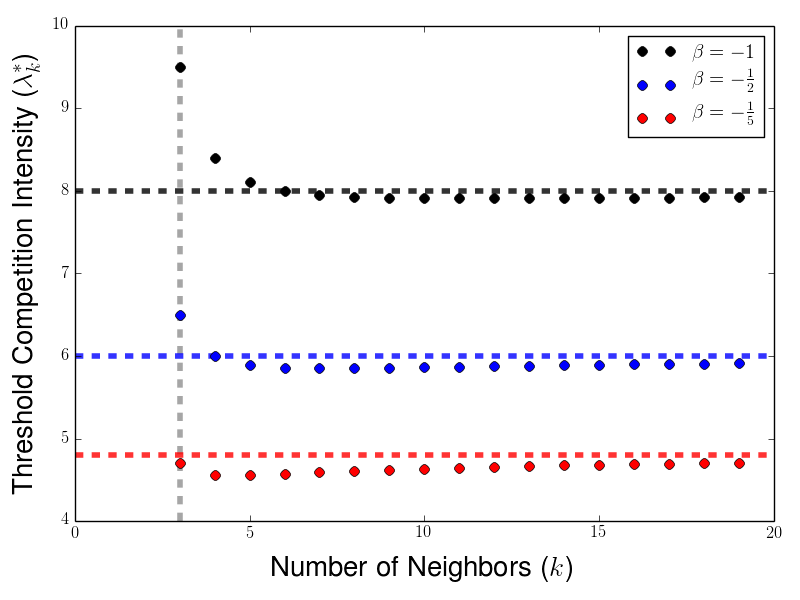}
      
     \caption[Threshold level of selection $\lambda^*_{k,DB}$ needed to achieve cooperation via multilevel selection as a function of neighborhood size $k$.]{The threshold level of selection $\lambda^*_{k,DB}$ needed to achieve cooperation via multilevel selection is a non-monotonic function of neighborhood size, $k$, for several different games. Threshold levels shown for $\beta = -1$ (black), $\beta = -\frac{1}{2}$, and $\beta = -\frac{1}{5}$, with other parameters fixed as $\gamma = \frac{3}{2}$, $\alpha = -1$, and $\theta = 2$. The dashed horizontal lines corresponds to the threshold $\lambda^*$ for multilevel selection for well-mixed interactions and within-group selection, with the color of the line corresponding to the value of $\beta$ for the dots of the same color. The vertical dashed line corresponds to the minimum neighborhood size $k=3$. For $\beta = -1$ and $\beta = -\frac{1}{2}$, there are sufficient sparse $k$-regular graphs such that $\lambda^*_k$ exceeds the well-mixed threshold, while the threshold is decreased for all finite neighborhood sizes $k$ when $\beta = -\frac{1}{5}$. }
     \label{fig:DBlambdathreshold}
 \end{figure}

In fact, we can fully characterize the class of PD games with $\alpha < 0$ for which the threshold selection strength $\lambda_{k}^{DB}$ to achieve cooperation is minimized by an intermediate neighborhood size $k$. In Proposition \ref{prop:klambdastarminDB}, we characterize the PD games for which a given neighborhood size $\hat{k}$ minimizes the threshold for achieving cooperation via multilevel selection $\lambda^*_{k,DB}$.

\begin{proposition} \label{prop:klambdastarminDB}
The neighborhood size $k$ satisfies $k = \argmin_{j \in \ZZ_{\geq 3}} \lambda^*_{k,DB}$ when the critical point $x^* = \frac{\gamma}{-2\alpha}$ of the group payoff function $G(x)$ under well-mixed interactions satisfies one of the following inequalities
\begin{equation}  \label{eq:xstarkthreshmininequality}
\left\{
     \begin{array}{cr}
       x^* > \frac{5}{7} + \frac{3}{7} \left( \frac{\beta}{\alpha} \right) & : k = 3\\
        B(k-1,k) > x^* > B(k,k+1) &: k \geq 4, 
     \end{array}
   \right.
\end{equation}
where 
\begin{equation} \label{eq:Bkj}
B(k,j) := \frac{(j+1) (k+1) }{2 (jk+2)} + \left( \frac{j+k+1}{jk+2} \right) \left( \frac{\beta}{\alpha} \right).
\end{equation}

\end{proposition}

For the PD game with $\beta = \alpha = -1$, we see the condition for the $3$-regular graph provide the lowest threshold $\lambda^*_{k,DB}$ is given by $x^* > \tfrac{5}{7} + \tfrac{3}{7} \left(1 \right) =  \frac{8}{7} > 1$. %
This means that, when $\beta = \alpha = -1$, all PD games with intermediate payoff optima experience a minimum threshold $\lambda^*_{k,DB}$ for graphs with neighborhood sizes $k > 3$. 

 In Figure \ref{fig:DBgrouppayoff}, we compare the average payoff at steady state for multilevel selection with DB updating on $k$-regular graphs (solid lines) with the baseline behavior of the model for the same game with game-theoretic interactions in well-mixed groups (dashed lines). We consider graphs with $k=3$ (left) and $k=5$ (right) neighbors, and consider three different values of $\beta$ in each case. We illustrate a game for which the collective outcome outperforms the well-mixed case for both neighborhood sizes ($\beta = -1/10$), as well as a game for which the collective outcome fares worse than the well-mixed case for both neighborhood sizes ($\beta = -1$). For a third game (with $\beta = -1/5$), we see that multilevel selection on a $5$-regular graph outperforms the well-mixed case, but placing interactions on a $3$-regular graph produces a worse collective outcome than achieved under well-mixed interactions.% and a game for which interactions on a $3$-regular graph produces a better outcome  %

  \begin{figure}[ht!]
     \centering
     \includegraphics[width = 0.48\textwidth]{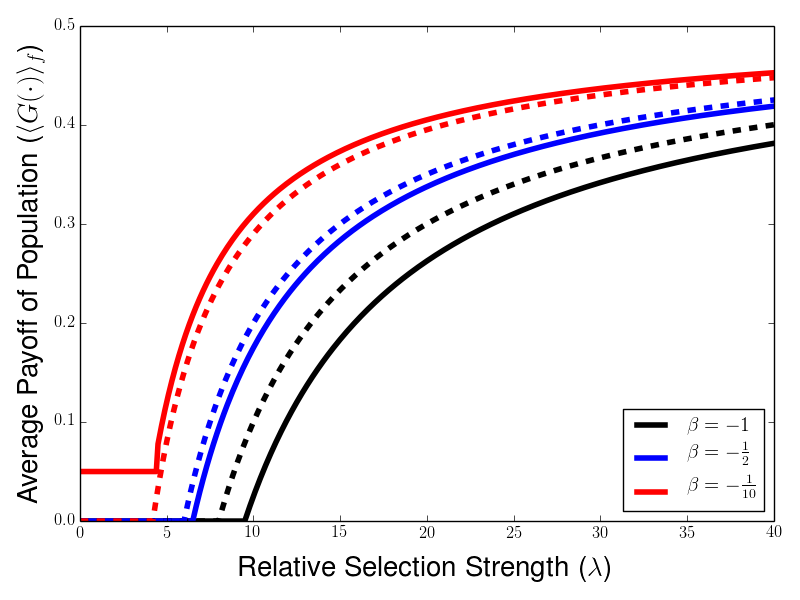}
      \includegraphics[width = 0.48\textwidth]{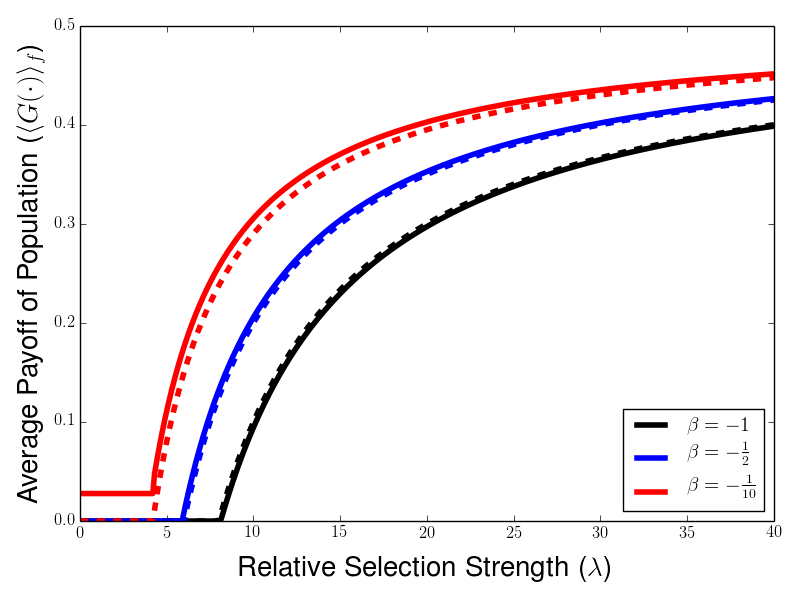}
     \caption[Average payoff of the steady state population for various values of $\lambda$, for game-theoretic interactions taking place on a $k$ -regular graph with DB rule or in a well-mixed group.]{Average steady state payoff with interactions on graphs can either increase or diminish relative to multilevel selection with well-mixed interactions. Average payoff of the steady state population for various values of relative selection strength $\lambda$, for game-theoretic interactions taking place on a $k$-regular graph (solid lines) or in a well-mixed group (dashed lines). Number of neighbors on graph is $k=3$ (left) or $k=5$ (right), and the value of $\beta$ varies between $\beta = -1$ (black lines), $\beta = -\frac{1}{2}$ (blue lines), and $\beta = -\frac{1}{10}$ (red lines). Other parameters chosen to be $\frac{3}{2}$, $\alpha = -1$, $\alpha = -1$, and $\theta = 2$.  }
     \label{fig:DBgrouppayoff}
 \end{figure}

\subsection{Birth-Death Updating} \label{sec:BD}

For birth-death updating, Ohtsuki and Nowak showed that full-defection remained globally stable for any $k$-regular graph for all of the games they considered \cite{ohtsuki2006replicator}. In fact, one can show that this is true for all PD games, and furthermore that the inclusion of graph-structured interactions with BD updating increases the threshold relative selection strength needed to achieve cooperation via multilevel selection relative to the case of well-mixed interactions.

To characterize the within-group dynamics under birth-death updating, we note that
\begin{equation}
    j_k^{BD}(x) = \beta + \alpha x + b_k^{BD} = \beta + \alpha x + \underbrace{\left( \frac{\alpha + 2 \beta}{k-2}\right)}_{< 0} < \beta + \alpha x < 0,
\end{equation}
where the last inequality holds for any PD game. As a consequence, we see from Equation \eqref{eq:graphwithingroupunnormalizedflux} that the fraction of defectors will always increase under the within-group dynamics under birth-death updating.  This means that placing interactions on a $k$-regular graph with birth-death updating cannot help achieve cooperation through individual-level selection alone. 

Furthermore, because the full-defection equilibrium is always stable under within-dynamics with BD updating, we know that we can use Equation \ref{eq:lambdastark} to express the threshold relative selection strength needed to produce density steady states supporting cooperation as 
\begin{equation*} \lambda^*_{k,BD} = \frac{-j_{k,BD} \theta}{G_k(1) - G_k(0)} = \frac{- \left( \beta + \alpha + \frac{\alpha + 2 \beta}{k-2} \right) \theta}{\gamma + \alpha} \\ > \frac{- \left( \beta + \alpha \right) \theta}{\gamma + \alpha} =: \lambda^*_{PD}.  \end{equation*}
$\lambda^*_{PD}$ is the threshold relative selection needed to support cooperation under well-mixed interactions for PD games, and therefore we see that a greater intensity of among-group selection is needed to facilitate cooperation via multilevel selection when interactions occur on $k$-regular graphs with the BD update rule. %
From these increased threshold levels, we see that BD updating does not promote cooperation at the individual level, but even makes achieving cooperation more difficult when groups are able to compete according to their collective payoffs.

\subsection{Imitation Updating} \label{sec:IM}

Now we discuss the dynamics of the IM update rule. This rule was shown in Ohtsuki and Nowak to have similar qualitative behavior to the DB update rule for the individual-level for several games, although having slightly weaker support for cooperation for a given game and neighborhood size $k$ \cite{ohtsuki2006replicator}. In Proposition \ref{prop:imthresholds}, we establish the game parameters $x^*$ and $\frac{\beta}{\alpha}$ for which within-group and multilevel selection are more capable of producing cooperation with interactions on $k$-regular graphs with IM updating relative to the well-mixed case. The proof of this is analogous to that of Proposition \ref{prop:dbthresholds}.

\begin{proposition} \label{prop:imthresholds}
Consider PD games with $\alpha < 0$ and within-group dynamics on a $k$-regular graph with IM updating. For such games, there exist the threshold quantities
\begin{subequations} \label{eq:imthresholds}
\begin{align}
    T_0^{IM}(k) & :=  \frac{k+3}{2k} + \left( \frac{k+1}{2} \right) \frac{\beta}{\alpha} \\
    T_1^{IM}(k) & := \frac{(k+3)(k-1)}{2k} + \left(\frac{k+1}{2} \right) \frac{\beta}{\alpha} \\
     T_{\lambda^*}^{IM}(k) & :=  \frac{k+3}{2k} + \left( \frac{3}{k} \right) \frac{\beta}{\alpha}
\end{align}
such that the full-defector equilibrium is unstable when $x^* > T_0^{IM}(k)$, the full-cooperator equilibrium is stable when $x^* > T_1^{IM}(k)$, and multilevel selection is favored for $k$-regular graph interactions relative to well-mixed interactions (i.e. $\lambda^*_{k,IM} < \lambda^*_{k \to \infty} := \frac{-(\beta + \alpha) \theta}{\gamma + \alpha}$) when $x^* > T_{\lambda^*}^{IM}(k)$. Furthermore, these threshold quantities satisfy the following ranking:
\begin{displaymath}
 T_{\lambda^*}^{IM}(k) \leq T_0^{IM}(k) \leq T_1^{IM}(k).
\end{displaymath}
\end{subequations}
\end{proposition}

 In Figure \ref{fig:IMparameter}, we illustrate, for the cases of $k=3$(left) and $k=4$(right), how the game parameters $x^*$ and $\frac{\beta}{\alpha}$ impact the different threshold levels for supporting cooperation provided in Proposition \ref{prop:imthresholds}. For the case of $k=3$, we see that $T_{\lambda^*}^{IM}(3) = 1 + 2 \left(\frac{\beta}{\alpha}\right) > 1$, and f we deduce from the rankings that no PD game with intermediate optima $x^* < 1$ can have cooperation improved relative to the well-mixed cases for IM updating with $k=3$. As shown in Figure \ref{fig:IMparameter}(left), all PD games with intermediate group payoff optima fall into the regime in which individual-level selection is not helped and multilevel selection is made more difficult by the addition of graph structure. For the $k=4$ case in Figure \ref{fig:IMparameter}, we see that there is a small portion of games with $x^* < 1$ with sufficiently small values of $\frac{\beta}{\alpha}$ such that cooperation can be increased both through individual level on graphs and through decreasing the threshold relative selection needed to achieve density steady states through multilevel selection.

 \begin{figure}[ht!]
     \centering
     \includegraphics[width = 0.48\textwidth]{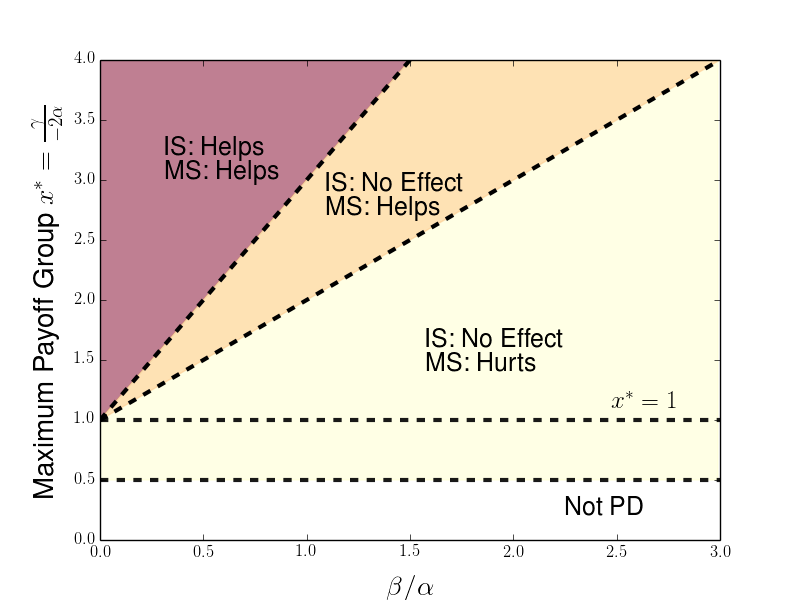}
      \includegraphics[width = 0.48\textwidth]{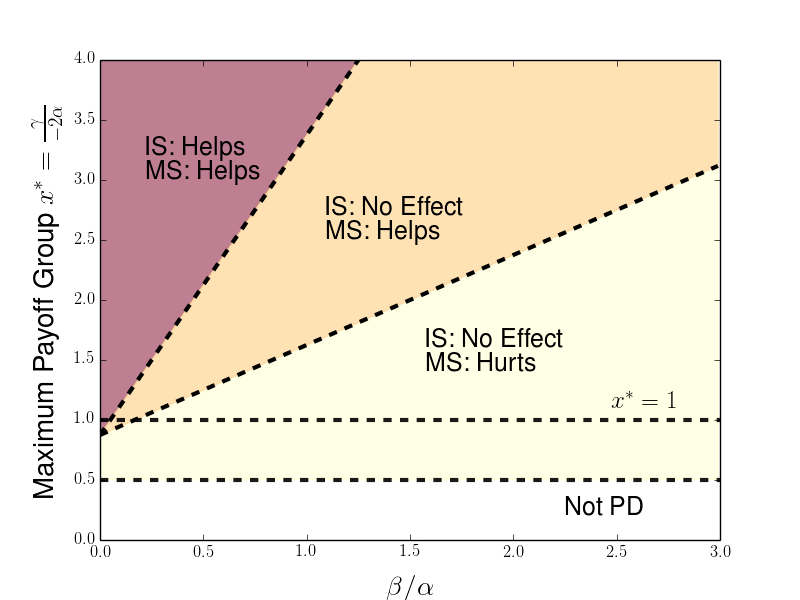}
     \caption[Regions of payoff parameter space in which multilevel selection and individual-level selection are helped or harmed by placing interactions on a $k$-regular graph with IM update rule, relative to well-mixed social interactions.]{Regions of payoff parameter space in which multilevel selection and individual-level selection are helped or harmed by placing interactions on a $k$-regular graph with the IM update rule, relative to well-mixed social interactions. Payoff matrices parametrized by Values of group optimum $x^*$ and the ratio $\frac{\beta}{\alpha}$, with graphs satisfying $k=3$ (left) and $k=4$. When $k=3$, all PD games with intermediate optima are harmed by within-group selection, while, when $k=4$, there are some games with intermediate optima in which within-group and among-group selection are improved relative to well-mixed interactions.}
     \label{fig:IMparameter}
 \end{figure}

Furthermore, we can prove an analogue of Proposition \ref{prop:klambdastarminDB} for the case of IM updating.
In particular, we find that the threshold for cooperation $\lambda^*_{k,IM}$ is minimized by neighborhood sizes $k > 3$ whenever $x^* < \tfrac{7}{6} + \tfrac{4}{3} \left( \tfrac{\beta}{\alpha} \right)$, meaning that $k$-regular graphs with more than three neighbors are more conducive to cooperation than $3$-regular graphs for all PD games with intermediate payoff optima.

 \subsection{Behavior in the Limit of Strong among-group Competition} \label{sec:graphlambdainfinity}

Finally, we consider how within-group network structure can change the level of cooperation achieved at steady state for relative selection strengths $\lambda \gg \lambda^*_{k,UR}$. In particular, because we saw from Example \ref{ex:alphaminus1} that PD games with intermediate payoff optima under well-mixed interactions can have group payoff functions $G_k(x)$ maximized by $x_k^* = 1$ for sufficiently small neighborhood sizes $k$, we can expect that placing interactions on a sufficiently sparse $k$-regular graph can help to erase the shadow of lower-level selection.

From \cite[Remark B.5]{cooney2022long}, we know that, for PD games with $\alpha < 0$, the steady state densities achieved by solutions of Equation \eqref{eq:graphpdegeneral} will concentrate, as $\lambda \to \infty$, upon a delta-function supported at the smallest fraction of cooperation at which $G_k(x)$ achieves the full-cooperator payoff $G_k(1) = \gamma + \alpha$. To determine this point of concentration for multilevel selection with interactions on graphs, we use Equation \eqref{eq:Gkparam} to see that the difference between $G_k(x)$ and $G_k(1)$ is given by
\begin{dmath} 
    G_k(x) - G_k(1) = {\left(\frac{k-2}{k-1} \right) \left(\gamma x + \alpha x^2 \right) + \left(\frac{1}{k-1} \right) \left(\gamma + \alpha \right) x - \left(\gamma + \alpha\right)} = \left( x - 1 \right) \left[ \alpha \left( \frac{k-2}{k-1} \right) + \left(\gamma + \alpha\right) \right].
\end{dmath}
As a result, the fraction of cooperation $\ol{x}_k$ at which the the steady state population concentrates is given by
\begin{equation} \label{eq:olxk}
    \ol{x}_k = \min\left\{ \left( \frac{k-1}{k-2} \right) \left( \frac{\gamma + \alpha}{-\alpha} \right), 1 \right\},
\end{equation}
and the concentration will occur at full-cooperation ($\ol{x}_k = 1$) for PD games with $\alpha < 0$ provided that
\begin{equation}
    k \leq 1 + \frac{\alpha}{\gamma + 2 \alpha}.
\end{equation}
By comparison, we can use Equation \eqref{eq:Gkparam} to see that the fraction of cooperators $x^*_k$ maximizing the average group payoff is given by
\begin{equation} \label{eq:xstark}
    x^*_k = \min\left\{\frac{\gamma}{-2 \alpha} +  \left( \frac{1}{k-2} \right) \left( \frac{\gamma + \alpha}{-\alpha} \right), 1 \right\},
\end{equation}
which is achieved by the full-cooperator group ($x^*_k$) provided that $k \leq 1 + \frac{\alpha}{\gamma + 2 \alpha}$.
 
 In Figure \ref{fig:IMkpeak}, we use our characterization of the optimal composition of cooperators $x^*_k$ from Equation \ref{eq:xstark} and the the group composition $\ol{x}_k$ at which steady state densities concentrate in the large $\lambda$ limit from Equation \ref{eq:olxk} to display the most-fit and modal group types at steady state for various numbers of neighbors $k$. For our choice of parameters, we see that the most-fit and most-abundant group types agree at full-cooperation between $k=3$ and $k=9$, while the most abundant group type at steady state decreases below the optimal level $x^*_k$ as the optimal level falls below full-cooperation for larger values of $k$. For even larger values of $k$, the most-fit and modal group types tend towards the values achieved by the underlying game in a population with well-mixed group interactions. In a sense, the behavior we see for $k$ between 3 and 9 is a means for eliminating the gap in modal and optimal levels of cooperation characteristic of the shadow of lower-level selection, as introducing graph-structured interactions changes the collective payoff scenario from one most favoring an intermediate level of cooperation to one pushing for as much cooperation as possible.

 \begin{figure}[H]
     \centering
     \includegraphics[width = 0.75\textwidth]{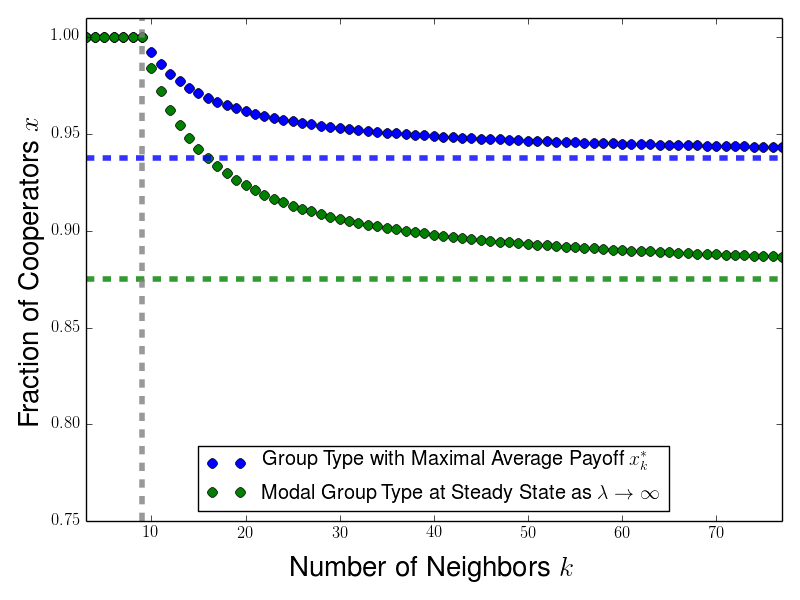}
      
     \caption[Comparison between group type with maximal payoff $x^*_{k}$and most abundant group type at steady state as $\lambda \to \infty$ for multilevel selection with interactions on $k$-regular graphs, plotted as a function of $k$.]{Comparison between group type with maximal payoff $x^*_{k}$ (blue) and most abundant group type at steady state in the limit as relative selection strength $\lambda \to \infty$ (green) for multilevel selection with the IM update rule, plotted as a function of the number of graph neighbors $k$. The blue and dashed lines correspond to the corresponding group type with maximal payoff and the group composition at which steady state densities concentration in the large-$\lambda$ limit for the case of well-mixed within-group interactions (or the $k\to \infty$ limit). For this example, we used $\gamma = \frac{31}{16}$ and $\alpha = -1$. The gray vertical dashed line corresponds to $k = 9$, the largest value of $k$ for which group payoff $G_k(x)$ is maximized by the full-cooperator group for our choice of parameters. }
     \label{fig:IMkpeak}
 \end{figure} 

\section{Threshold Selection Strength in Finite Population Model of Multilevel Selection} \label{sec:finite}

Having explored the dynamics of multilevel selection in our PDE model obtained in the limit of infinitely many groups, each with infinitely many members,  %that the shadow of lower-level selection is a generic feature of the two-level replicator equation for the PD and HD games,
it is also useful to explore whether behaviors in the continuum limit of a two-level birth-death process have an analogue or a signature in the underlying stochastic process. In particular, we are curious about the extent to which the threshold formula $\lambda^*$ arising as a tug-of-war between the individual incentive to defect and the collective incentive to cooperate can be seen in a corresponding finite population model. In this section, We will consider a a slight modification of the finite population model introduced by Traulsen and coauthors \cite{traulsen2006evolution,traulsen2008analytical}, in which there are $m$ groups each comprised of $n$ individuals, and consider the special case in which the within-group events take place on a much faster timescale than the events corresponding to among-group competition. In the finite population setting, the population will eventually either fix to a state of all full-cooperator groups or of all full-defector groups, so our main quantities of interest will be the fixation probability of a single cooperator (resp. defector) in a population otherwise consisting of defectors (resp. defectors). Here, we will modify the framing of the Traulsen-Nowak model \cite{traulsen2008analytical} by parametrizing the selection strength of birth events at the individual and group levels $w_I$ and $w_G$, and study how the relative selection intensity $W := w_G / w_I$ affects the relative fixation probability of cooperators and defectors. This approach will allow us to compare results for relative fixation in this stochastic setting to the quantities introduction in Section \ref{sec:model} for the corresponding PDE model of multilevel selection. 

We assume that each individual plays the game against every member of their group, and assume self-interactions for simplicity. The average payoff in these $n$ interactions for a cooperator and defector in a group with $i$ cooperators is given by
\begin{subequations} \begin{align} \pi_C(i) &= \frac{i}{n} R + \left(1 - \frac{i}{n} \right) S  \\ \pi_D(i) &= \frac{i}{n} T + \left( 1 - \frac{i}{n} \right) P \end{align} \end{subequations}
Here we will use an exponential mapping of payoff to reproduction or imitation probability, which was introduced by Traulsen and coauthors \cite{traulsen2008analytical} to study fixation for multilevel selection, and has shown to provide analytical characterizations of relative fixation probabilities in terms of the relevant parameters of population structure \cite{cooney2016assortment,tarnita2009strategy}. 
For births and deaths of individuals, we describe how the number of cooperators $i$ evolves according to a discrete-time Moran process. In such a process, one individual is chosen to give birth per time-step, and the offspring replaces a randomly chosen member of the group, perhaps including its parent. In an $i$-cooperator group, we choose to model the fecundity or ability to reproduce of a cooperator $F^C_i$ and defector $F^D_i$ by the exponential function of payoffs
\begin{subequations} \begin{align} F^C_i &= \exp\left(w_I \pi_C(i) \right) \\ F^D_i &= \exp\left(w_I \pi_D(i) \right)   \end{align} \end{subequations}
In a time-step, one individual is chosen to reproduce with probability proportional to its fecundity and its offspring replaces a randomly chosen member of the parent's group (including possibly its parent). The probability of a transition from an $i$-cooperator group to an $i'$ cooperator group, denoted by $p_{i,i'}$ is given by 
\begin{subequations}
\begin{align} p_{i,i+1} &= \label{eq:newcooperator} \left( \frac{i F^C_i}{i F^C_i + (n-i) F^D_i}\right) \left( 1 - \frac{i}{n} \right) \\ p_{i,i-1} &= \label{eq:newdefector} \left( \frac{(n-i) F^D_i}{i F^C_i + (n-i) F^D_i} \right) \left( \frac{i}{n} \right) \\ p_{i,i} &= 1 - p_{i,i-1} - p_{i,i+1} \\ p_{i,i'} &= 0 \: \: \textnormal{for $i' \not\in \{i-1,i,i+1\}$}  \end{align} \end{subequations}
where the first terms in Equation \ref{eq:newcooperator} (resp. \ref{eq:newdefector}) correspond to the birth of a cooperator (resp. defector) based on payoff and the death of a randomly chosen defector (resp. cooperator). Denoting by $\rho^I_C$  as the probability of fixation of a single cooperator mutant within an otherwise all-defector group and $\rho^I_D$ as the corresponding fixation probability of a single defector, we have that the ratio of fixation probabilities is given by
\begin{equation} \label{eq:individualfixationratio} \frac{\rho^I_C}{\rho^I_D} = \ds\prod_{i=i}^{n-1} \frac{p_{i,i+1}}{p_{i,i-1}}  = \ds\prod_{i=i}^{n-1} \frac{F_i^C}{F_i^D} =  \ds\prod_{i=i}^{n-1} \frac{\exp\left(w_I \pi_C(i) \right)}{\exp\left(w_I \pi_D(i) \right)} = \prod_{i=i}^{n-1}  \exp\left(w_I \left[ \pi_C(i) - \pi_D(i) \right] \right) \end{equation}
\cite{nowak2006evolutionary,traulsen2008analytical}, and we use a useful property of the exponential payoff-to-fitness mapping to show that 
\begin{align}  \frac{\rho^I_C}{\rho^I_D} &= \exp\left(w_I \ds\sum_{i=1}^{n-1}  \left[ \pi_C(i) - \pi_D(i)\right]  \right) \nonumber \\ &= \exp\left( w_I \ds\sum_{i=1}^{n-1} \left[ \left( R - S - T + P \right) \left( \frac{i}{n} \right) + \left( S - P\right)   \right]\right) \nonumber \\ &= \exp\left(w_I (n-1) \left[\frac{R+S-T-P}{2} \right] \right)  \label{eq:individualfixationratiosimpler}\end{align}

As in the model of Traulsen and coauthors, we assume that there is a separation of timescales between within-group and among-group dynamics, in which individual-level dynamics fix cooperation or defection within a group before selection has time to act upon among-group replication events \cite{traulsen2008analytical}. 

Then we can study the competition among groups on a slower time scale, which we model as a group-level Moran process in which full-cooperator groups and full-defector groups give birth with probability proportional to average payoff, replacing randomly chosen groups. We describe the population state by the number of groups $j$ of the $m$ groups composed of cooperators, and characterize group-level fecundity for full-cooperator and full-defector groups as 
\begin{subequations} \label{eq:groupfecundity} \begin{align} H^C &= \exp\left(w_G \pi_C(n) \right) = \exp\left(w_G R \right) \\   H^D &= \exp\left(w_G \pi_D(0) \right)  =  \exp\left(w_G P \right)   \end{align} \end{subequations}
where $w_G$ corresponds to a selection intensity or sensitivity to payoff differences for among-group reproduction events. With the same reasoning as in the individual level dynamics, the probability of transitions $q_{j,j'}$ between a state with $j$ full-cooperator groups to $j'$ full-cooperator groups are given by
\begin{subequations} \label{eq:grouptransitionprob}
\begin{align} q_{j,j+1} &= \label{eq:newcooperatorgroup} \left( \frac{j H^C}{j H^C + (m-j) H^D}\right) \left( 1 - \frac{j}{m} \right) \\ q_{j,j-1} &= \label{eq:newdefectorgroup} \left( \frac{(m-j) H^D}{j H^C + (m-j) H^D} \right) \left( \frac{j}{m} \right) \\ q_{j,j} &= 1 - q_{j,j-1} - q_{j,j+1} \\ q_{j,j'} &= 0 \: \: \textnormal{for $j' \not\in \{j-1,j,j+1\}$}  \end{align} \end{subequations}
Denoting by $\rho^G_C$ and $\rho^G_D$ the probabilities of a fixation of a single full-cooperator group and of a single full-defector group, we can similarly use Equations \ref{eq:groupfecundity} and \ref{eq:grouptransitionprob} to find that
\begin{equation} \label{eq:groupfixationratio} \frac{\rho_C^G}{\rho_D^G} = \ds\prod_{j=1}^{m-1} \frac{q_{j,j+1}}{q_{j,j-1}} = \ds\prod_{j=1}^{m-1} \frac{H^C}{H^D} = \ds\prod_{j=1}^{m-1} \exp\left( w_G \left[ R - P \right] \right) = \exp \left( w_G \left(m-2\right) \left[ R - P \right] \right) \end{equation}
Combining the fixation of both a cooperator within its own group and the subsequent fixation of the ensuing full-cooperator group on the time scale of among-group compeitition tells us that fixation probability of a single cooperator in the group-structured population is given by $\rho_C = \rho_C^I \rho_C^G$ \cite{traulsen2008analytical}. Similarly noting that the fixation probability of a single defector is given by $\rho_D = \rho_D^I \rho_D^G$, we have that the ratio of fixation probabilities for a single cooperator and a single defector is given by 
\begin{equation} \label{eq:rhoCrhoD} \frac{\rho_C}{\rho_D} = \exp\left(w_G \left(m-2 \right) \left( R - P \right) -  \frac{w_I (n-1)}{2}  \left( R + S - T - P \right) \right) \end{equation}
We see that the fixation of a single cooperator is favored over fixation of a single defector ($\rho_C > \rho_D$) when 
\begin{equation} \label{eq:wGwI} W = \frac{w_G}{w_I} >  \frac{(n-1)}{2 (m-2)} \left( \frac{T + P - R - S}{R - P} \right) \end{equation}
Thinking as we do in the PDE limit about a fixed payoff matrix and changing the relative selection strength $w_G / w_I$, we have a critical ratio of selection strengths 
\begin{equation} \label{eq:Wstar} W^*:=  \frac{n-1}{2 (m-2)} \left( \frac{T + P - R - S}{R - P} \right)\end{equation} such that fixation of cooperators is favored when $w_G > W^* w_I$ and fixation of defectors is favored when $w_G < W^* w_I$. This threshold value $W^*$ can be rewritten in terms of the parameters $\alpha$, $\beta$, and $\gamma$ as 
\begin{equation} \label{eq:Wstarparam} W^* = \frac{-(n-1)}{2 (m-2)} \left( \frac{\alpha + 2 \beta}{\gamma + \alpha} \right) \end{equation}
Further, recalling that $\pi_D(x) - \pi_C(x) = - \left( \beta + \alpha x\right)$ and that $G(x) = \gamma x + \alpha x^2$, we see that we can write $W^*$ in terms of these two functions evaluated at the endpoints $0$ and $1$ as 
\begin{equation} \label{eq:Wstarpayoff} W^* = \frac{n-1}{2 (m-2)} \left( \frac{\left( \pi_D(1) - \pi_C(1)\right) + \left(\pi_D(0) - \pi_C(0)\right)}{G(1) - G(0)} \right) \end{equation}
In particular, this means that the fixation probability of defectors is improved by increasing the defector's payoff advantage in otherwise full-cooperator groups or in otherwise full-defector groups, while fixation of cooperation is favored by improving the payoff of full-cooperator groups or hurting the payoff of full-defector groups. %

 For the PD game, we know that $\alpha + 2 \beta = \left(R - T\right) \left(S-P\right)  < 0$, which tells us that the minimum threshold $W^* > 0$, so fixation to cooperation can only be favored over fixation to defection if among-group competition is sufficiently strong. %
 Notably, approaching the edge case for the PD in which $R \to P$, or equivalently $\gamma \to - \alpha$, we have that $G(1) = \gamma + \alpha \to 0 = G(0)$, so the threshold $W^* \to \infty$ as $R \to P$. Therefore, we require infinitely large among-group selection strength $w_G$ to favor fixation of a cooperator over fixation of a defector in the limit as $R \to P$, and therefore a mutant defector will have the fixation advantage for any finite $w_I$ and $w_G$. The behavior of $W^*$ in this limit is analogous to the behavior in the PDE limit of the multilevel dynamics, in which the threshold relative selection strength $\lambda^* \to \infty$ as $R \to P$ (and correspondingly as $G(1) \to G(0)$, and therefore the dominance of the defectors' within-group payoff advantage causes the convergence of the population to a delta-function at full-defector groups for any finite relative level of selection strength $\lambda$ in this limit. 
 
 The phenomenon in which $W^* \to \infty$ as $R \to P$ for the PD game can be thought of as somewhat analogous to the behavior by which $\lambda^* \to \infty$ as $G(1) \to G(0)$ for the infinite population PD dynamics discussed in Section \ref{sec:model}. As we saw in the deterministic limit, we can understand the expression of Equation \ref{eq:Wstarpayoff} for $W^*$ as a balance found from the tug-of-war between the group advantage of greater payoffs of full-cooperator groups over full-defector groups, expressed as $G(1) - G(0)$, and the individual payoff advantage of being a defector rather than a cooperator in either of these groups,  encoded through $\pi_D(0) - \pi_C(0)$ and $\pi_D(1) - \pi_D(1)$. In the finite population case, this arises due to the fact that populations composed entirely of full-cooperator groups or entirely of full-defector groups are absorbing states of the stochastic process. This tug-of-war between within-group and among-group incentives at the endpoint still holds in the PDE model in which within-group and among-group competition occur on comparable timescales, suggesting how survival of cooperation via multilevel selection relies heavily of the ability of full-cooperator groups to achieve sufficient success to overcome within-group pressures for the dominance of defection.  %
 
 The formulas obtained for the survival or fixation of cooperation in the stochastic and deterministic settings may motivate further extensions. One direction for future work is to extend the analysis of our nested stochastic process to explore asymptotic expressions for fixation and switching probabilities between monomorphic equilibria \cite{mcloone2018stochasticity,deville2017finite,park2017extinction}, as well as considering fixation probabilities and stationary distributions in finite populations for games with more than two strategies in the absence or presence of mutations, respectively  \cite{fudenberg2006imitation,vasconcelos2017stochastic,ferreira2020fixation}. In additional, one can explore the role that spatial among-group structure can play in facilitating or hindering the evolution of cooperation via multilevel selection. Akdeniz and van Veelen has suggested that localized spatial competition between group may hinder the spread of cooperation, showing that placing groups on a cycle graph with with $k$-nearest neighbor among-group competition hurts the relative fixation probability of cooperation \cite{akdeniz2020cancellation}. Further exploration of multilevel competition with a range of among-group graph structures may allow us to see whether there are group-level analogues to amplifiers or suppressors of selection seen in individual-level selection \cite{adlam2015amplifiers,lieberman2005evolutionary,hindersin2015most,hindersin2016should}. Using the within-group structure coefficient and the extent to which among-group structure amplifies or suppresses selection, we can understand how population structure within groups and among groups can impact the ability to produce cooperation via multilevel selection.

\end{document}